\documentclass[a4paper,11pt]{article}
\pdfoutput=1 % if your are submitting a pdflatex (i.e. if you have
\usepackage[utf8x]{inputenc}
\usepackage[T1]{fontenc}
\usepackage{tikz}
\usetikzlibrary{matrix,arrows}
\usepackage{enumitem}
\usepackage{mwe}
\usepackage[colorlinks=true
  ,urlcolor=blue
  ,anchorcolor=blue
  ,citecolor=blue
  ,filecolor=blue
  ,linkcolor=blue
  ,menucolor=blue
  ,linktocpage=true
  ,pdfproducer=medialab
  ,pdfa=true]{hyperref}
\usepackage{amsmath}
\usepackage{amsthm}
\usepackage{amssymb}
\usepackage{amsfonts}
\usepackage{mathrsfs}
\usepackage{bbm}
\usepackage{abstract}
\usepackage{graphicx}
\usepackage{placeins}
\usepackage{subfigure}
\usepackage{nicefrac}
\usepackage{booktabs}
\usepackage{bookmark}
\usepackage{multirow}
\usepackage{dsfont}
\usepackage{longtable}
\usepackage{braket}
\usepackage{tikz}
\usepackage{scalerel}
\usepackage{listings}
\usepackage{color}
\usepackage[all]{xy}
\usepackage{tensor}
\usepackage{faktor}
\usepackage{lipsum}
\theoremstyle{definition}
\newtheorem{definition}{Definition}[section] 
 
\theoremstyle{definition}
\newtheorem{remark}[definition]{Remark}

\theoremstyle{definition}

\theoremstyle{plain}

\theoremstyle{plain}

\theoremstyle{plain}

\theoremstyle{plain}

\usepackage{jheppub} % for details on the use of the package, please
\title{\boldmath Holst-MacDowell-Mansouri action for (extended) supergravity with boundaries and super Chern-Simons theory}

\author[1]{K. Eder,}
\author[2]{H. Sahlmann,}
\affiliation[]{Friedrich-Alexander-Universität Erlangen-Nürnberg (FAU),\\
Institute for Quantum Gravity (IQG),\\Staudtstr. 7,D-91058 Erlangen, Germany}
\emailAdd{konstantin.eder@gravity.fau.de}
\emailAdd{hanno.sahlmann@gravity.fau.de}
\abstract{In this article, the Cartan geometric approach toward (extended) supergravity in the presence of boundaries will be discussed. In  particular, based on new developments in this field, we will derive the Holst variant of the MacDowell-Mansouri action for $\mathcal{N}=1$ and $\mathcal{N}=2$ pure AdS supergravity in $D=4$ for arbitrary Barbero-Immirzi parameters. This action turns out to play a crucial role in context of boundaries in the framework of supergravity if one imposes supersymmetry invariance at the boundary. For the $\mathcal{N}=2$ case, it follows that this amounts to the introduction of a $\theta$-topological term to the Yang-Mills sector which explicitly depends on the Barbero-Immirzi parameter. This shows the close connection between this parameter and the $\theta$-ambiguity of gauge theory.\\
We will also discuss the chiral limit of the theory, which turns out to 
possess some very special properties such as the manifest invariance of the resulting action under an enlarged gauge symmetry. Moreover, we will show that demanding supersymmetry invariance at the boundary yields a unique boundary term corresponding to a super Chern-Simons theory with $\mathrm{OSp}(\mathcal{N}|2)$ gauge group. In this context, we will also derive boundary conditions that couple boundary and bulk degrees of freedom and show equivalence to the results found in the D'Auria-Fré approach in context of the non-chiral theory. These results provide a step towards of quantum description of supersymmetric black holes in the framework of loop quantum gravity.}
\begin{document} 
\maketitle
\flushbottom

\section{Introduction}\label{superCartan geometry}
The physics of boundaries, in particular the interaction between degrees of freedom on the boundary and those in the bulk play an important role in diverse areas of physics, from solid state physics to gravity. In the latter area, this is particularly the case for the horizons of black holes. Bekenstein-Hawking entropy \cite{Bekenstein:1973ur,Bardeen:1973gs} assigns the black hole an entropy as if there was one bit of information encoded in each Planck unit of its horizon area, and Hawking radiation looks as if it was perfectly thermal at its surface \cite{Hawking:1974sw}. The holographic principle as advocated by 't Hooft, Susskind and others states that the entire state of the black hole is represented on its surface \cite{Susskind:1994vu}.  In loop quantum gravity, a picture that is consistent with these holographic ideas emerges partially from an observation about the classical theory and its boundary at the horizon: If a spacetime with an inner boundary is considered, and boundary conditions are imposed at the inner boundary consistent with it being an apparent horizon, the symplectic structure attains a contribution corresponding to a Chern-Simons theory on the horizon \cite{Smolin:1995vq,Ashtekar:1999yj,Ashtekar:2000eq,Kaul:2000kf,Engle:2010kt}. In the quantum theory, the excitations of the gravitational field create defects in the horizon Chern-Simons theory, thereby changing the size of the state space and accounting for the black hole entropy \cite{Rovelli:1996dv,Ashtekar:1997yu,Domagala:2004jt,Engle:2009vc,Agullo:2010zz}. %\cite{} These results are for Einstein gravity, but progress has been made to quantize supergravity with loop quantum gravity methods.\cite{} It would be very interesting to extend the consideration of black hole entropy to supersymmetric black holes to make contact with results from string theory.  The present work in that it describes the classical theory induced by $D=4$ supergravity on an inner boundary. 

Boundary theories also play a crucial role in string theory such as in context of the celebrated $\mathrm{AdS}_{d+1}/\mathrm{CFT}_d$ conjecture \cite{Maldacena:1997re,Witten:1998qj,Aharony:1999ti}, a far reaching duality which attracted a lot of interest since its discovery by Maldacena. It describes a duality between string theory on a $d+1$ dimensional asymptotically anti-de Sitter spacetime and a $d$-dimensional conformal field theory on the boundary such as, most prominently, between type IIB superstring theory on an $\mathrm{AdS}_5\times S^5$ background and $\mathcal{N}=4$ super Yang-Mills theory living on the boundary. In the low-energy limit of string theory -- aka supergravity -- this holographic correspondence has been studied very intensively. There, one observes a one-to-one correspondence between fields of the bulk supergravity theory satisfying certain boundary conditions and quantum operators associated to the boundary conformal field theory.
%Finally, this duality describes a relation between a strongly coupled quantum field theory on the boundary to a weakly-coupled gravity theory of one dimension higher. As a consequence, it also has the potential to provide a lot of insights on the non-perturbative regime of gauge theories, a regime which one has only limited access to and are therefore not that well-understood.\\

On the other hand, boundaries in string theory have also recently been explored in \cite{Mikhaylov:2014aoa} where a specific brane configuration in the framework of type IIB superstring theory has been considered consisting of a stack of D3 branes on two sides of a NS5 brane where the worldvolume theory on the D3 branes corresponds to a maximally supersymmetrc Yang-Mills theory with $\mathrm{U}(n)$ gauge group. There, it has been observed that the boundary theory is described by a super Chern-Simons theory with gauge group given by the super unitary group $\mathrm{U}(m|n)$ and complex Chern-Simons level. But also other configurations have been considered leading to super Chern-Simons theories with gauge supergroups $\mathrm{OSp}(m|n)$.
% For instance, in ??, boundaries have been discussed imposing Dirichlet boundary conditions on the fields. However, it has been argued in ??, that imposing local boundary conditions on all the fields of supergravity may break local supersymmetry. Hence, in contrast, in ??, a different approach has been considered studying a more general class of (not necessarily local) boundary conditions that also maintain supersymmetry invariance at the boundary.

In the context of supergravity, there exist various different approaches on the proper description of boundaries (see e.g. \cite{Gibbons:1976ue,Avramidi:1997hy,vanNieuwenhuizen:2005kg,Belyaev:2007bg}). More recently, boundaries in supergravity have been considered in \cite{Andrianopoli:2014aqa,Andrianopoli:2020zbl,Concha:2018ywv,Ipinza:2016con,Concha:2014tca} in the framework of the so-called \emph{geometric approach} also commonly known as the D'Auria-Fré approach \cite{DAuria:1982uck,Castellani:1991et}. There, a systematic approach for $D=4$ pure supergravity theories both with and without a cosmological constant has been developed, by studying the most general class of possible boundary terms that are compatible with the symmetry of the bulk Lagrangian. By demanding supersymmetry invariance at the boundary, these boundary terms then turned out to be determined even uniquely. Moreover, within this formalism, one finds in both cases, i.e., with and without a cosmological constant, that the associated boundary conditions are not of Dirichlet-type but require the vanishing of the supercurvatures on the boundary. Finally, it follows that the resulting action of the theory including bulk and boundary degrees of freedom takes a very intriguing form which, for $\mathcal{N}=1$ and nontrivial cosmological constant, exactly reproduces the well-known MacDowell-Mansouri action \cite{MacDowell:1977jt}. In particular, in this way, a similar structure has been found for $\mathcal{N}=1$, $D=4$ flat supergravity \cite{Concha:2018ywv} as well as $\mathcal{N}=2$, $D=4$ pure AdS supergravity \cite{Andrianopoli:2014aqa}.

LQG is a program of quantum gravity originially based on canonical quantization of variables introduced by Sen, Ashtekar, Immirzi and Barbero \cite{Sen:1982qb,Ashtekar:1986yd,Barbero:1994ap,Immirzi:1996di} for Einstein gravity. These variables have the remarkable property that they embed the phase space of gravity in that of Yang-Mills theory. It was pointed out in \cite{Holst:1995pc} that all these variables can be obtained from an action that differs from the Palatini action by a certain topological term defined by an operator on the Lie algebra of the structure group. This modification of the gravitational action is thus one of the foundations of the theory.

While LQG is much less ambitious then string theory in terms of unification, it has very interesting results to its credit,  such as a kinematical representation that carries a unitary representation of spatial diffeomorphisms, quantized spatial geometry, an account of black hole entropy as described above in more detail, and a path integral formulation in terms of so-called spin foams. Also in the latter, the Holst action plays an important role. In fact, it was found that the additional term is essential to obtain viable gravitational amplitudes \cite{Alesci:2007tx,Alesci:2007tg,Alesci:2008ff}. 

The present work is part of a long running effort to allow the quantization techniques of LQG to be applied to supergravity. We will describe the status quo below. In this article we want to study the classical theory of boundaries in supergravity in $D=4$ for actions that contain a modification analogous to the one that leads to Sen/Ashtekar/Immirzi/Barbero variables \cite{Sen:1982qb,Ashtekar:1986yd,Barbero:1994ap,Immirzi:1996di}. There are several reasons why this is an interesting topic to study. To begin with, it is indispensable to understand action and canonical theory in the presence of generic boundaries before studying the situation for (apparent) horizons of black holes. This will in turn shed further light on the quantum description of black holes in loop quantum gravity (LQG), allow for the treatment of supersymmetric black holes with the same methods, and a comparison with string theory results. It is relevant in other contexts in which boundaries play a role, such as the ones listed in the beginning of this introduction, in particular also in the discussion of subsystems in gravity \cite{Donnelly:2016auv}. 
Finally, in this work we develop and use a concise and geometrical form of the supergravity actions with Holst-type modifications, which may be useful in itself for future work.

In context of LQG,  $\mathcal{N}=1$ supergravity in terms of self-dual variables has been studied e.g. in \cite{Jacobson:1987cj,Fulop:1993wi}. In particular, in \cite{Fulop:1993wi}, on the kinematical level, a hidden $\mathfrak{osp}(1|2)$ gauge symmetry in the constraint algebra has been observed which subsequently has been used to formulate a quantum theory à la LQG by Gambini, Pullin et al. \cite{Gambini:1995db} and Ling and Smolin \cite{Ling:1999gn} and in context of spin foam models in $D=3$ e.g. in \cite{Livine:2003hn,Livine:2007dx}. Extended $\mathcal{N}=2$, $D=4$ chiral supergravity has been studied e.g. in \cite{Sano:1992jw,Tsuda:2000er}, and in terms of a constrained super BF-theory in \cite{Ezawa:1995nj}. Finally, these considerations have also been extended to include real variables in \cite{Sawaguchi:2001wi,Tsuda:1999bg,Eder:2020uff} as well as Bodendorfer et al. \cite{Bodendorfer:2011hs,Bodendorfer:2011pb,Bodendorfer:2011pc} in context of higher dimensional SUGRA theories.
Finally boundaries in supergravity in the framework of LQG have been discussed using self-dual variables already a long time ago in \cite{Ling:2000ss,Ling:2003yw}. Interestingly, there the authors already seem to suggest that topological terms contained in the (chiral) MacDowell-Mansouri action may play a role in (quantum) description of boundaries in supergravity. 

In what follows, we want to study this question from a more general perspective following newer developments in the geometric approach \cite{Andrianopoli:2014aqa,Andrianopoli:2020zbl} and pointing out the importance of supersymmetry invariance at the boundary and also explicitly including real Ashtekar-Barbero variables. Moreover, we will start from the original supergravity Lagrangians instead from constrained field theories. We will not use the formalism of isolated or dynamical horizons (\cite{Ashtekar:2000sz,Ashtekar:2001jb,Bodendorfer:2013jba} and \cite{Ashtekar:2004cn} for an overview and further literature) as it has not yet been thoroughly studied in the context of supergravity, and because its boundary conditions seem not to be well-adapted to the requirement of local supersymmetry at the boundary. Rather, following \cite{Andrianopoli:2014aqa,Andrianopoli:2020zbl} we make the condition of local supersymmetry extending to the boundary the guiding principle for finding appropriate boundary terms and boundary conditions. 

Therefore, using the interpretation of supergravity in terms of a super Cartan geometry, we will derive the Holst modification of the MacDowell-Mansouri action for arbitrary Barbero-Immirzi parameter $\beta$ for $\mathcal{N}=1$ and $\mathcal{N}=2$ pure AdS supergravity as derived in \cite{Andrianopoli:2014aqa} which as mentioned above, by construction, already contains the most general class of boundary terms maintaining supersymmetry invariance at the boundary. To this end, inspired by \cite{Randono:2006ru,Wise:2009fu,Freidel:2005ak} in context of ordinary gravity, we will introduce some kind of a $\beta$-deformed inner product induced by a $\beta$-dependent quasi-projection operator $\mathbf{P}_{\beta}$ defined on super Lie algebra-valued differential forms. As we will see, this approach then also allows for a very elegant and unified discussion of the chiral limit of the theory. There, it follows that $\mathcal{P}_{\pm i}$ becomes a projection operator onto a proper subalgebra of the super anti-de Sitter algebra corresponding to the (complex) orthosymplectic group $\mathrm{OSp}(\mathcal{N}|2)_{\mathbb{C}}$. As a consequence, the resulting action becomes manifestly invariant under $\mathrm{OSp}(\mathcal{N}|2)_{\mathbb{C}}$ thus revealing the underlying enlarged gauge symmetry of the chiral theory for both cases in a very clear way.

In particular, it follows that the resulting boundary terms correspond to a super Chern-Simons action with gauge group given by the supergroups $\mathrm{OSp}(\mathcal{N}|2)_{\mathbb{C}}$ and complex Chern-Simons level. For $\mathcal{N}=1$, we will also prove explicitly that the full action is indeed invariant under left- and right-handed supersymmetry transformations on the boundary and even turns out to be fixed uniquely by this requirement. Finally, we will derive the boundary conditions of the full theory describing the coupling between the bulk and boundary degrees of freedom. As we will see, these turn out to be in strong analogy to the standard boundary conditions as usually applied in LQG and, in partcular, transform covariantly under the enlarged gauge symmetry of the chiral theory.

The structure of the paper is as follows. At the beginning, we recall very briefly some basic elements of the Cartan geometric approach to $\mathcal{N}=1$ pure AdS supergravity and discuss the most general class of possible boundary terms following \cite{Andrianopoli:2014aqa,Andrianopoli:2020zbl}. We then define in section \ref{Holst N=1} the Holst-MacDowell-Mansouri action by introducing a $\beta$-dependent quasi projecion operator. In section \ref{sec:N1chiral}, we will disucss the chiral limit of the theory, derive a manifestly $\mathrm{OSp}(1|2)$-gauge invariant form of the action and  discuss the resulting boundary theory. We then repeat this procedure for the $\mathcal{N}=2$ extended case in the subsequent sections \ref{Section:Pure N=2} and \ref{Holst N=2}. In particular, we will extend the $\mathcal{N}=2$ pure supergravity action as found in \cite{Andrianopoli:2014aqa} to arbitrary $\beta$ including the most general class of boundary terms compatible with local supersymmetry. We will then discuss the chiral limit as well as the boundary theory in section \ref{Section:Chiral N=2} and compare our results with those found in \cite{Andrianopoli:2014aqa,Andrianopoli:2020zbl}.\\
\\
We work in signature $(-+++)$. The gravitational coupling constant is denoted by $\kappa=8 \pi G$, the cosmological constant by $\Lambda=-3/L^2$. Indices $I,J\ldots=0,\ldots,3$ are local Lorentz indices. 4D Majorana spinor indices are denoted by $\alpha,\beta,\ldots$ whereas $A,B,\ldots$ and $A',B',\ldots$ are left- and right-handed Weyl spinor indices, respectively. Finally, $R$-symmetry indices are denoted by $p,q,r,s,\ldots$. The conventions for the gamma matrices as well as spinor calculus can be found in appendix \ref{Appendix:Gamma}.
\newpage

% D'Auria-Fré approach \cite{DAuria:1982uck,Castellani:1991et}. \\
%-Townsend full classification in D=3
%-Livine and Oeckl: D=3 spin foams and fermions
%-Adrianopoli: Kerr black hole and Chern-Simons theory
%-LQG: previous results in context of supergravity
%-Mikaylov-Witten: string theory and super Chern-Simons theory

\section{Geometric $\mathcal{N}=1$ supergravity with boundaries}\label{N=1 SUGRA with inner bdy}
In this section, we want to briefly recall the geometric interpretation of $\mathcal{N}=1$ AdS-supergravity\footnote{The four-dimensional anti-de Sitter spacetime is an embedded submanifold of the semi-Riemannian manifold $\mathbb{R}^{2,3}$ equipped with the metric $\eta_{AB}=\mathrm{diag}(-+++-)$, that is
\begin{equation*}
\mathrm{AdS}_4:=\{x\in\mathbb{R}^{2,3}|\,\eta_{AB}x^Ax^B=-L^2\}
%\label{eq:N1action}
\end{equation*}
with $L$ the so-called anti-de Sitter radius} in terms of a super Cartan geometry as discussed in more detail in \cite{Eder:2020erq,Eder:2021ans}. For a review of the relevant supergroups and our choice of conventions see appendix \ref{Appendix:Supergroups}. Moreover, following \cite{Andrianopoli:2014aqa}, we discuss the extension of the theory in the presence of boundaries and the implementation of supersymmetric boundary conditions.\\
Pure $D=4,\,\mathcal{N}=1$ AdS supergravity can be described in terms of a super Cartan geometry modeled on the flat super Klein geometry ($\mathrm{OSp}(1|4)$,$\mathrm{Spin}^+(1,3)$) corresponding to super anti-de Sitter space.\footnote{In order to properly include anticommutative fermionic degrees of freedom, super Cartan geometries have to be defined on enriched categories, see \cite{Eder:2020erq,Eder:2021ans} for more details.} Using the decomposition $\mathfrak{osp}(1|4)=\mathbb{R}^{1,3}\oplus\mathfrak{spin}^+(1,3)\oplus\Delta_{\mathbb{R}}$ of the super Lie algebra, with odd part $\Delta_{\mathbb{R}}$ corresponding to a real Majorana representation, the super Cartan connection $\mathcal{A}$ of the theory takes the form 
\begin{equation}
\mathcal{A}=e^IP_I+\frac{1}{2}\omega^{IJ}M_{IJ}+\psi^{\alpha}Q_{\alpha}
\label{eq:1.1}
\end{equation}
with $e_I$ the co-frame, $\omega^{IJ}$ the spin connection and $\psi^{\alpha}$ the Rarita-Schwinger field. Moreover, the horizontal forms contained in the Cartan connection build up the \emph{supervielbein} or \emph{super soldering form} $E=e^IP_I+\psi^{\alpha}Q_{\alpha}$ which provides a local identification of the curved supermanifold with the flat model given by super $\mathrm{AdS}_4$. This is a direct consequence of so-called \emph{(super) Cartan condition}. The action of the theory takes the form
\begin{align}
S(\mathcal{A})=:\frac{L^2}{\kappa}\int_M\mathscr{L}:=&\frac{1}{2\kappa}\int_{M}\left(\frac{1}{2}F(\omega)^{IJ}\wedge e^K\wedge e^L\epsilon_{IJKL}+
i\bar{\psi}\wedge\gamma_{*}\gamma_ID^{(\omega)}\psi\wedge e^I\right.\nonumber\\
&\left.-\frac{1}{4L}\bar{\psi}\wedge\gamma^{IJ}\psi\wedge e^K\wedge e^L\epsilon_{IJKL}+\frac{1}{4L^2}e^I\wedge e^J\wedge e^K\wedge e^L\epsilon_{IJKL}  \right)
\label{eq:N1action}
\end{align}
with $F(\omega)^{IJ}=\mathrm{d}\omega^{IJ}+\tensor{\omega}{^I_K}\wedge\tensor{\omega}{^{KJ}}$ the curvature of the spin-connection $\omega$ and $D^{(\omega)}\psi=\mathrm{d}\psi+\frac{1}{4}\omega^{IJ}\gamma_{IJ}\wedge\psi$ the induced exterior covariant derivative. As explained in detail in \cite{Eder:2020erq,Eder:2021ans}, the underlying supersymmetry of the theory can be described using the bijective correspondence between Cartan connections and Ehresmann connections, i.e. ordinary gauge fields playing a role for instance in Yang-Mills gauge theories, on the associated $\mathrm{OSp}(1|4)$-bundle. One can then interpret local supersymmetry in terms of local gauge transformations in the odd direction of the supergroup. That this is possible is a particularity of the pure $\mathcal{N}=1$ case since, generically, SUSY transformations have to be regarded as superdiffeomorphisms. For the pure $\mathcal{N}=1$ case, the appropriate superdiffeomorphisms turn out to take the same form as super gauge transformations. Thus, using the (graded) commutation relations \eqref{eq:C8a}-\eqref{eq:C8d}, it follows that, under such kind of gauge transformations, the individual fields transform as
\begin{align}
\delta_{\epsilon}e^I=\frac{1}{2}\bar{\epsilon}\gamma^I\psi,\quad\delta_{\epsilon}\psi^{\alpha}&=D^{(\omega)}\epsilon^{\alpha}-\frac{1}{2L}e^I\tensor{(\gamma_I)}{^{\alpha}_{\beta}}\wedge\epsilon^{\beta},\quad\delta_{\epsilon}\omega^{IJ}=\frac{1}{2L}\bar{\epsilon}\gamma^{IJ}\psi
\label{eq:1.3}
\end{align}
for some Grassmann odd Majorana spinor $\epsilon^{\alpha}$. It then turns out that \eqref{eq:N1action} is indeed invariant under \eqref{eq:1.3} provided the spin-connection satisfies its field equations.

So far, we have excluded the possibility of boundaries in our discussion. In the presence of boundaries, it follows that one needs to add additional (topological) boundary terms in order to maintain functional differentiability of the action functional \eqref{eq:N1action}. Moreover, in case of supergravity, one is particular is interested in boundary terms which also ensure invariance of the full action under supersymmetry transformations at the boundary. It turns out that this requirement strongly restricts the structure of possible boundary terms to be added to the theory. Therefore, following \cite{Andrianopoli:2014aqa}, one notes that the only possible topological terms which are consistent with the symmetries of the bulk Lagrangian $\mathscr{L}\equiv\mathscr{L}_{\text{bulk}}$ in \eqref{eq:N1action} are of the form
\begin{equation}
\mathscr{L}_{\text{bdy}}=C_1F(\omega)^{IJ}\wedge F(\omega)^{KL}\epsilon_{IJKL}+C_2\left(D^{(\omega)}\bar{\psi}\wedge\gamma_{*}D^{(\omega)}\psi+\frac{i}{8}\epsilon_{IJKL}F(\omega)^{IJ}\wedge\bar{\psi}\wedge\gamma^{KL}\psi\right)
\label{eq:1.4}
\end{equation}   
for any constant coefficients $C_1,C_2$. The first term in \eqref{eq:1.4} is the well-known \emph{Gauss-Bonnet term} which is indeed topological. The second term can equivalently be written as
\begin{equation}
D^{(\omega)}\bar{\psi}\wedge\gamma_{*}D^{(\omega)}\psi+\frac{i}{8}\epsilon_{IJKL}F(\omega)^{IJ}\wedge\bar{\psi}\wedge\gamma^{KL}\psi=\mathrm{d}(\bar{\psi}\wedge\gamma_{*}D^{(\omega)}\psi)
\label{eq:1.5}
\end{equation}
and therefore is also a total derivative. As shown in \cite{Andrianopoli:2014aqa}, if one requires invariance of the full Lagrangian $\mathscr{L}_\text{full}=\mathscr{L}_{\text{bulk}}+\mathscr{L}_{\text{bdy}}$ under supersymmetry transformations, then the coefficients $C_1,C_2$ are uniquely fixed to particular values given by $C_1=1/8$ and $C_2=i/(2L)$, respectively. We will in fact show this explicitly in the chiral theory in section \ref{sec:N1chiral} below. Moreover, using \eqref{eq:1.5} it follows that the full Lagrangian $\mathscr{L}_{\text{full}}$ is that of the MacDowell-Mansouri action \cite{MacDowell:1977jt,Castellani:2013iq,Andrianopoli:2014aqa}, i.e. quadratic in the super Cartan curvature $F(\mathcal{A})$ to be defined in the next section such that
\begin{equation}
\mathscr{L}_{\text{full}}=\frac{1}{4}F(\mathcal{A})^{IJ}\wedge F(\mathcal{A})^{KL}\epsilon_{IJKL}+\frac{i}{L}F(\mathcal{A})^{\alpha}\wedge F(\mathcal{A})^{\delta}(C\gamma_{*})_{\alpha\delta}
\end{equation}
 To summarize, the boundary terms for $D=4$, $\mathcal{N}=1$ AdS-supergravity in the presence of boundaries are uniquely fixed by requirement of supersymmetry invariance at the boundary, and they are neatly contained in MacDowell-Mansouri action. 
\subsection{Holst-MacDowell-Mansouri action of $\mathcal{N}=1$ SUGRA}\label{Holst N=1}
In this section, we want to discuss $D=4$, $\mathcal{N}=1$ AdS-supergravity in the context of LQG. As in the previous section, we want to explicitly the include the possibility of boundaries in the theory. We therefore need to derive a Holst variant of the MacDowell-Mansouri action for arbitrary \emph{Barbero-Immirzi parameter} $\beta$ which is either assumed to be real, i.e.,  $\beta\in\mathbb{R}^{\times}=\mathbb{R}\textbackslash\{0\}$, in case of real variables, or purely imaginary, i.e., $\beta=\pm i$ in case of the chiral theory to be discussed in detail in the next section.\\
A derivation of the Holst action of $D=4$, $\mathcal{N}=1$ supergravity via a MacDowell-Mansouri action, by adding a suitable topological term and  treating $\beta$ as kind of a $\theta$-ambiguity similar to Yang-Mills theory, has been given in \cite{Obregon:2012zz} and for the special case of the chiral theory in \cite{Nieto:1996wn}. Here, we want to follow the ideas of \cite{Randono:2006ru,Wise:2009fu} in the context of classical first order Einstein gravity and its reformulation in terms of constrained BF-theory \cite{Freidel:2005ak}. As we will show, these ideas can naturally be extended to supergravity by introducing a \emph{$\beta$-deformed inner product} on the superalgebra.\\
To this end, note that, using the explicit representation of $\mathfrak{osp}(1|4)$ as summarized in appendix \ref{Appendix:Supergroups},  the generators of $\mathfrak{spin}^+(1,3)$ take the form $M_{IJ}=\frac{1}{2}\gamma_{IJ}$. One can then define an operator $\mathcal{P}_{\beta}$ on $\mathfrak{spin}^+(1,3)$ via
\begin{equation}
\mathcal{P}_{\beta}:=\frac{\mathds{1}+ i\beta\gamma_*}{2\beta}:\,\mathfrak{spin}^+(1,3)\rightarrow\mathfrak{spin}^+(1,3).
\label{eq:3.1}
\end{equation}
That this operator indeed leaves $\mathfrak{spin}^+(1,3)$ invariant follows from $i\gamma_*\gamma_{IJ}=\frac{1}{2}\tensor{\epsilon}{_{IJ}^{KL}}\gamma_{KL}$ which, moreover, yields the important identity
\begin{equation}
\mathcal{P}_{\beta}\gamma^{IJ}=i\gamma_*\tensor{{P_{\beta}}}{^{IJ}_{KL}}\gamma^{KL},\,\text{with}\quad\tensor{{P_{\beta}}}{^{IJ}_{KL}}:=\frac{1}{2}\left(\delta^I_{[K}\delta^J_{L]}-\frac{1}{2\beta}\tensor{\epsilon}{^{IJ}_{KL}}\right)
\label{eq:3.2}
\end{equation}
Since the odd part of $\mathfrak{osp}(1|4)$, in particular, defines a Clifford module, we can naturally extend $\mathcal{P}_{\beta}$ to an operator on $\Delta_{\mathbb{R}}$ via $\mathcal{P}_{\beta}Q_{\alpha}=Q_{\delta}\tensor{(\mathcal{P}_{\beta})}{^{\delta}_{\alpha}}$. Hence, using the identification $\mathfrak{osp}(1|4)\cong\mathbb{R}^{1,3}\oplus\mathfrak{spin}^+(1,3)\oplus\Delta_{\mathbb{R}}$, one can introduce an operator $\mathbf{P}_{\beta}$ on the super Lie algebra (or rather its complexification), by setting
\begin{equation}
\mathbf{P}_{\beta}:=\underline{\mathbf{0}}\oplus\mathcal{P}_{\beta}\oplus\mathcal{P}_{\beta}:\,\mathfrak{osp}(1|4)\rightarrow\mathfrak{osp}(1|4).
\label{eq:3.3}
\end{equation}
Using this operator, we can define an inner product on the super Lie algebra. Note first that a standard Adjoint-invariant inner product on $\mathfrak{osp}(1|4)$ is given by the supertrace $\braket{\cdot,\cdot}:=\mathrm{str}$. When combined with \eqref{eq:3.3}, this yields a the corresponding $\beta$-deformed inner product setting
\begin{equation}
\braket{X,Y}_{\beta}:=\mathrm{str}(X\cdot\mathbf{P}_{\beta}Y),\quad\forall X,Y\in\mathfrak{osp}(1|4)
\label{eq:3.4}
\end{equation}
which is invariant under $\mathrm{Spin}^+(1,3)$ but not under the full supergoup $\mathrm{OSp}(1|4)$. Making use of this inner product, we can now formulate the Holst-MacDowell-Mansouri action of $\mathcal{N}=1$, $D=4$ supergravity. It is given by  
\begin{equation}
S_{H-MM}(\mathcal{A})=\frac{L^2}{\kappa}\int_{M}{\braket{F(\mathcal{A})\wedge F(\mathcal{A})}_{\beta}},
\label{eq:3.5}
\end{equation}
where $F(\mathcal{A})$ is the \emph{Cartan curvature} of $\mathcal{A}$ defined as 
\begin{equation}
F(\mathcal{A})=\mathrm{d}\mathcal{A}+\frac{1}{2}[\mathcal{A}\wedge\mathcal{A}]=\mathrm{d}\mathcal{A}+\frac{1}{2}(-1)^{|T_{\underline{A}}||T_{\underline{B}}|}\mathcal{A}^{\underline{A}}\wedge\mathcal{A}^{\underline{B}}\otimes[T_{\underline{A}},T_{\underline{B}}]
\label{eq:3.6}
\end{equation}
with respect to the homogeneous basis $(T_{\underline{A}})_{\underline{A}}$ of $\mathfrak{osp}(1|4)$, $\underline{A}\in(I,IJ,\alpha)$, where the minus sign in (\ref{eq:3.5}) appears due to the (anti)commutation of $T_{\underline{A}}$ and $\mathcal{A}^{\underline{B}}$. Using the graded commutation relations \eqref{eq:C8a}-\eqref{eq:C8d} in case $\mathcal{N}=1$ as well as $[M_{IJ},P_K]=\eta_{IK}P_J-\eta_{JK}P_I$, it follows that the translational and Lorentzian sub components of $F(\mathcal{A})$ take the form 
\begin{align}
F(\mathcal{A})^I=\Theta^{(\omega) I}-\frac{1}{4}\bar{\psi}\wedge\gamma^I\psi,\quad F(\mathcal{A})^{IJ}=F(\omega)^{IJ}+\frac{1}{L^2}\Sigma^{IJ}-\frac{1}{4L}\bar{\psi}\wedge\gamma^{IJ}\psi
\label{eq:3.7}
\end{align}
respectively, with $\Sigma^{IJ}:=e^I\wedge e^J$ and $\Theta^{(\omega)I}=\mathrm{d}e^I+\tensor{\omega}{^I_J}\wedge e^J$ the \emph{torsion 2-form} associated to $\omega$. For the odd part of the curvature, we find
\begin{align}
F(\mathcal{A})^{\alpha}=D^{(\omega)}\psi^{\alpha}-\frac{1}{2L}e^I\wedge\tensor{(\gamma_{I})}{^{\alpha}_{\beta}}\psi^{\beta}.
\label{eq:3.8}
\end{align}
To see that this in fact leads to the Holst action of $\mathcal{N}=1$ AdS supergravity, let us expand the action \eqref{eq:3.5}. If we use \eqref{eq:3.7} and \eqref{eq:3.8}, we find\footnote{Here and in the following, $\bar{\psi}\wedge\gamma^{\bullet\bullet}\psi$ denotes a $\mathfrak{spin}^+(1,3)$-valued 2-form with components $\bar{\psi}\wedge\gamma^{IJ}\psi$.}
\begin{align}
\braket{F(\mathcal{A})\wedge F(\mathcal{A})}_{\beta}=&\frac{1}{4}F(\mathcal{A})^{IJ}\wedge F(\mathcal{A})^{KL}\braket{M_{IJ},M_{KL}}_{\beta}-F(\mathcal{A})^{\alpha}\wedge F(\mathcal{A})^{\delta}\braket{Q_{\alpha},Q_{\delta}}_{\beta}\nonumber\\
=&\braket{F(\omega)\wedge F(\omega)}_{\beta}+\frac{2}{L^2}\braket{\Sigma\wedge F(\omega)}_{\beta}-\frac{1}{2L}\braket{F(\omega)\wedge\bar{\psi}\wedge\gamma^{\bullet\bullet}\psi}_{\beta}\nonumber\\
&-\frac{1}{2L^3}\braket{\Sigma\wedge\bar{\psi}\wedge\gamma^{\bullet\bullet}\psi}_{\beta}+\frac{1}{L^4}\braket{\Sigma\wedge\Sigma}_{\beta}+\frac{1}{L}F(\mathcal{A})^{\alpha}\wedge F(\mathcal{A})^{\delta}(C\mathcal{P}_{\beta})_{\alpha\delta},
\label{eq:3.9}
\end{align}
where we used that $\braket{Q_{\alpha},Q_{\delta}}_{\beta}=-\frac{1}{L}(C\mathcal{P}_{\beta})_{\alpha\delta}$ which can be checked by direct computation using the explicit representation given in appendix \ref{Appendix:Supergroups}. Using $\braket{M_{IJ},M_{KL}}_{\beta}=-\frac{i}{4}\tensor{{P_{\beta}}}{^{MN}_{KL}}\times\mathrm{tr}(\gamma_{IJ}\gamma_{MN}\gamma_{*})=\tensor{{P_{\beta}}}{^{MN}_{KL}}\epsilon_{IJMN}$, this yields
\begin{align}
&\braket{F(\omega)\wedge F(\omega)}_{\beta}+\frac{2}{L^2}\braket{\Sigma\wedge F(\omega)}_{\beta}-\frac{1}{2L}\braket{F(\omega)\wedge\bar{\psi}\wedge\gamma^{\bullet\bullet}\psi}_{\beta}-\frac{1}{2L^3}\braket{\Sigma\wedge\bar{\psi}\wedge\gamma^{\bullet\bullet}\psi}_{\beta}\nonumber\\
&+\frac{1}{L^4}\braket{\Sigma\wedge\Sigma}_{\beta}\nonumber\\
=&\braket{F(\omega)\wedge F(\omega)}_{\beta}+\frac{2}{L^2}\braket{\Sigma\wedge F(\omega)}_{\beta}-\frac{1}{8L}F(\omega)^{IJ}\wedge\tensor{{P_{\beta}}}{^{KL}_{MN}}\bar{\psi}\wedge\gamma^{MN}\psi\epsilon_{IJKL}\nonumber\\
&-\frac{1}{8L^3}\Sigma^{IJ}\wedge\tensor{{P_{\beta}}}{^{KL}_{MN}}\bar{\psi}\wedge\gamma^{MN}\psi\epsilon_{IJKL}+\frac{1}{4L^4}\Sigma^{IJ}\wedge\tensor{{P_{\beta}}}{^{KL}_{MN}}\Sigma^{MN}\epsilon_{IJKL}\nonumber\\
=&\braket{F(\omega)\wedge F(\omega)}_{\beta}+\frac{2}{L^2}\braket{\Sigma\wedge F(\omega)}_{\beta}+\frac{i}{8L}F(\omega)^{IJ}\wedge\bar{\psi}\wedge\gamma_{*}\gamma^{KL}\mathcal{P}_{\beta}\psi\epsilon_{IJKL}\nonumber\\
&+\frac{i}{8L^3}\Sigma^{IJ}\wedge\bar{\psi}\wedge\gamma_{*}\gamma^{KL}\mathcal{P}_{\beta}\psi\epsilon_{IJKL}+\frac{1}{8L^4}\Sigma^{IJ}\wedge\Sigma^{KL}\epsilon_{IJKL}.
\label{eq:3.10}
\end{align}
On the other hand, we have
\begin{align}
\frac{1}{L}F(\mathcal{A})^{\alpha}\wedge F(\mathcal{A})^{\delta}(C\mathcal{P}_{\beta})_{\alpha\delta}=&\braket{D^{(\omega)}\psi\wedge D^{(\omega)}\psi}_{\beta}-\frac{1}{L^2}\bar{\psi}\wedge\boldsymbol{\gamma}\wedge\mathcal{P}_{\beta}D^{(\omega)}\psi\nonumber\\
&-\frac{1}{4L^3}\bar{\psi}\wedge\gamma_{IJ}\mathcal{P}_{-\beta}\psi\wedge\Sigma^{IJ}.
\label{eq:3.11}
\end{align}
where we set $\boldsymbol{\gamma}:=e^I\gamma_I$. Adding \eqref{eq:3.10} and \eqref{eq:3.11} and again using the identity $\tensor{\epsilon}{_{IJ}^{KL}}\gamma_{KL}=2i\gamma_{IJ}\gamma_{*}$ which yields
\begin{align}
&\frac{i}{8L^3}\Sigma^{IJ}\wedge\bar{\psi}\wedge\gamma_{*}\gamma^{KL}\mathcal{P}_{\beta}\psi\epsilon_{IJKL}-\frac{1}{4L^3}\bar{\psi}\wedge\gamma_{IJ}\mathcal{P}_{\beta}\psi\wedge\Sigma^{IJ}\nonumber\\
=&\frac{i}{8L^3}\bar{\psi}\wedge\gamma_{*}\gamma^{KL}\big[\mathcal{P}_{\beta}+\mathcal{P}_{-\beta}\big]\psi\wedge\Sigma^{IJ}\epsilon_{IJKL}=-\frac{1}{8L^3}\bar{\psi}\wedge\gamma^{KL}\psi\wedge\Sigma^{IJ}\epsilon_{IJKL}
\label{eq:3.12}
\end{align}
it follows that the Holst-MacDowell-Mansouri action takes the form
\begin{align}
S_{H-MM}(\mathcal{A})=\frac{1}{2\kappa}\int_{M}&\Sigma^{IJ}\wedge(P_{\beta}\circ F(\omega))^{KL}\epsilon_{IJKL}-\bar{\psi}\wedge\boldsymbol{\gamma}\wedge\frac{\mathds{1}+i\beta\gamma_*}{\beta} D^{(\omega)}\psi\nonumber\\
-&\frac{1}{4L}\bar{\psi}\wedge\gamma^{KL}\psi\wedge\Sigma^{IJ}\epsilon_{IJKL}+\frac{1}{4L^2}\Sigma^{IJ}\wedge\Sigma^{KL}\epsilon_{IJKL}+S_{\text{bdy}}
\end{align}
where we wrote $(P_{\beta}\circ F(\omega))^{IJ}=\tensor{{P_{\beta}}}{^{IJ}_{KL}}F(\omega)^{KL}$. Moreover, $S_{\text{bdy}}$ denotes a boundary term given by
\begin{align}
S_{\text{bdy}}(\mathcal{A})=\frac{L^2}{\kappa}\int_{M}&\braket{F(\omega)\wedge F(\omega)}_{\beta}+\braket{D^{(\omega)}\psi\wedge D^{(\omega)}\psi}_{\beta}-\frac{1}{4L}F(\omega)^{IJ}\wedge\bar{\psi}\wedge\gamma_{IJ}\mathcal{P}_{\beta}\psi
\label{eq:3.13.1}
\end{align}
Thus, we see, up to a topological term, \eqref{eq:3.5} indeed reduces to the Holst action of $\mathcal{N}=1$, $D=4$ AdS supergravity \cite{Obregon:2012zz} and which in the limit of a vanishing cosmological constant yields the respective action of Poincaré supergravtiy \cite{Tsuda:1999bg,Kaul:2007gz,Eder:2020uff}. To see that \eqref{eq:3.13.1} is in fact purely topological, note that, by the \emph{Bianchi-identity}, we have
\begin{equation}
D^{(\omega)}D^{(\omega)}\psi=\kappa_{\mathbb{R}*}(F(\omega))\wedge\psi=\frac{1}{4}F(\omega)^{IJ}\gamma_{IJ}\wedge\psi
\label{eq:3.13}
\end{equation}
such that by the $\mathrm{Spin}^+(1,3)$-invariance of the inner product, this yields
\begin{align}
\braket{D^{(\omega)}\psi\wedge D^{(\omega)}\psi}_{\beta}&=\mathrm{d}\!\braket{\psi\wedge D^{(\omega)}\psi}_{\beta}+\braket{\psi\wedge D^{(\omega)}D^{(\omega)}\psi}_{\beta}\nonumber\\
&=\mathrm{d}\!\braket{\psi\wedge D^{(\omega)}\psi}_{\beta}+\frac{1}{4L}F(\omega)^{IJ}\wedge\bar{\psi}\wedge\gamma_{IJ}\psi
\end{align}
Moreover, according to the general discussion in the appendix \ref{Chern-Simons} in case of arbitrary (super) connections, we have
\begin{align}
\braket{F(\omega)\wedge F(\omega)}_{\beta}=\mathrm{d}\!\braket{\omega\wedge\mathrm{d}\omega+\frac{1}{3}\omega\wedge[\omega\wedge\omega]}_{\beta}
\end{align}
Thus, to summarize, it follows that \eqref{eq:3.13.1} can be equivalently be written in the form
\begin{align}
S_{\text{bdy}}(\mathcal{A})=\frac{L^2}{\kappa}\int_{\partial M}\braket{\omega\wedge\mathrm{d}\omega+\frac{1}{3}\omega\wedge[\omega\wedge\omega]}_{\beta}+\braket{\psi\wedge D^{(\omega)}\psi}_{\beta}
\label{eq:3.14}
\end{align}
and hence, in particular, is non-vanishing in the presence of boundaries. According to the general discussion in the previous section, this is the most general boundary term one can have in the context of $\mathcal{N}=1$, $D=4$ anti-de Sitter supergravity in the framework of LQG if one requires invariance of the full theory under local supersymmetry transformations.

Therefore, note that the deformed action \eqref{eq:3.5} is invariant under the same SUSY transformations \eqref{eq:1.3} as in the standard theory. In fact, since in the $\mathcal{N}=1$ case the SUSY transformations can be regarded as super gauge transformations, it follows that the transformation of the super Cartan curvature takes the form 
\begin{equation}
    \delta_{\epsilon}F(\mathcal{A})=-[\epsilon,F(\mathcal{A})]
\end{equation}
Thus, if we set $\rho^{\alpha}=F(\mathcal{A})^{\alpha}$ this implies that the variation of the Lagrangian in \eqref{eq:3.5} yields
\begin{align}
    \delta_{\epsilon}\mathscr{L}=&-\braket{[\epsilon,F(\mathcal{A})]\wedge F(\mathcal{A})}_{\beta}-\braket{F(\mathcal{A})\wedge[\epsilon,F(\mathcal{A})]}_{\beta}\nonumber\\
    =&\frac{1}{4L}F(\mathcal{A})^{IJ}\wedge\bar{\epsilon}\gamma^{KL}\rho\,\tensor{{P_{\beta}}}{^{MN}_{KL}}\epsilon_{IJMN}-\frac{1}{L^2}F(\mathcal{A})^I\wedge\bar{\rho}\mathcal{P}_{\beta}\gamma_I\epsilon\nonumber\\
    &-\frac{1}{2L}F(\mathcal{A})^{IJ}\wedge\bar{\epsilon}\mathcal{P}_{\beta}\gamma_{IJ}\rho=-\frac{1}{L^2}F(\mathcal{A})^I\wedge\bar{\rho}\mathcal{P}_{\beta}\gamma_I\epsilon
\end{align}
Hence, it follows that the Lagrangian of the full $\beta$-deformed action is invariant under the SUSY transformations, both in the bulk and at the boundary, provided that the supertorsion constraint $F(\mathcal{A})^I=0$ is satisfied which is equivalent to requiring that the spin connection $\omega$ satisfies its equations of motion. As far as the bulk theory is concerned, this was actually to be expected since, as will be proven explicitly for the case $\mathcal{N}=2$ in section \ref{Holst N=2} below, at second order, the deformed action coincides with the standard action up to topological terms, so that the SUSY variations are indeed unaltered. 

Due to the deformed inner product appearing in \eqref{eq:3.14} the boundary action contains additional topological terms compared to the standard theory. For instance, writing out the bosonic contribution in \eqref{eq:3.14}, this yields
\begin{align}
\braket{F(\omega)\wedge F(\omega)}_{\beta}=\frac{1}{8}F(\omega)^{IJ}\wedge F(\omega)^{KL}\epsilon_{IJKL}+\frac{1}{4\beta}F(\omega)^{IJ}\wedge F(\omega)_{IJ}
\end{align}
so that the bosonic part of the boundary action splits into the ordinary Gauss-Bonnet term as in the standard theory as well as an additional topological \emph{Pontryagin term}. As discussed in \cite{Perez:2017cmj}, these are in fact the most general boundary terms one can expect in the pure bosonic theory in case of a finite Barbero-Immirzi parameter which are compatible with the symmetries of the Lagrangian. The fermionic contribution in \eqref{eq:3.14} takes the form $\braket{\psi\wedge D^{(\omega)}\psi}_{\beta}=\frac{1}{L}\bar{\psi}\wedge\mathcal{P}_{\beta} D^{(\omega)}\psi$. Similar to the bulk theory as discussed e.g. in \cite{Eder:2020uff,Tsuda:1999bg}, in the canonical description of the boundary theory, the operator $\mathcal{P}_{\beta}$ implies that the covariant derivative can be re-rexpressed in terms of the covariant derivative associated to the real \emph{Ashtekar-Barbero connection} 
\begin{equation}
A^{\beta}:=\Gamma+\beta K    
\end{equation}
which is a $\mathrm{SU}(2)$ connection. These are the standard variables used for the canonical quantization of the theory in the framework of LQG.

Finally, comparing with \eqref{eq:A11}, one may suspect that the boundary action \eqref{eq:3.14} almost looks like a super Chern-Simons action. Note, however, that the deformed inner product is not invariant under the full group $\mathrm{OSp}(1|4)$ but only under the action of its bosonic subgroup so that \eqref{eq:3.14}, at least in general, will not correspond to a Chern-Simons action with a supergroup as a gauge group. This is of course in contrast e.g. to the IH formalism \cite{Ashtekar:2004cn}, where the IH boundary conditions imply that the boundary theory is generically described in terms of a Chern-Simons theory. However, as we will see in the next section, this changes drastically in case of the chiral theory.

\subsection{The chiral theory: Chiral Palatini action and super Chern-Simons theory on the boundary}
\label{sec:N1chiral}
Having derived the most general form of the Holst action of AdS supergravity in the presence of boundaries which also incorporates local supersymmetry invariance, we now want to focus on the special case of a imaginary Barbero-Immirzi parmareter $\beta=\pm i$. As we will, the resulting theory has many interesting properties and in fact leads to numerous intriguing structures which seem to be well-compatible with the underlying supersymmetry.\\
Therefore, in what follows, let us set $\beta=-i$ (the other case can be treated in complete analogy). In this case, the operator \eqref{eq:3.3} takes the form $\mathcal{P}_{\beta=-i}=i\frac{\mathds{1}+\gamma_*}{2}$ so that $\mathbf{P}_{\beta=-i}=:i\mathbf{P}^+$ where, according to the general discussion in appendix \ref{Appendix:Supergroups} (see also \cite{Eder:2020erq}
 for more details), $\mathbf{P}^+$ defines a projection 
\begin{equation}
\mathbf{P}^+:\,\mathfrak{osp}(1|4)_{\mathbb{C}}\rightarrow\mathfrak{osp}(1|2)_{\mathbb{C}}
\label{eq:4.1}
\end{equation}
onto a proper sub superalgebra of $\mathfrak{osp}(1|4)$ given by the (complex) orthosymplectic algebra $\mathfrak{osp}(1|2)_{\mathbb{C}}$ corresponding to the superalgebra of $\mathcal{N}=1$, $D=2$ super anti-de Sitter space (in fact, it turns out that \eqref{eq:4.1} even defines a morphism of superalgebras). It then follows that inner product \eqref{eq:3.4} reduces to the standard inner product $\braket{\cdot,\cdot}$ on $\mathfrak{osp}(1|2)_{\mathbb{C}}$ given by the supertrace which, in particular, is invariant under the Adjoint representation of $\mathrm{OSp}(1|2)_{\mathbb{C}}$. Applying the projection \eqref{eq:4.1} on the super Cartan connection \eqref{eq:1.1}, this yields the \emph{super Ashtekar connection} 
\begin{equation}
\mathcal{A}^+:=\mathbf{P}^+\mathcal{A}=A^{+ i}T^+_i+\psi^A Q_{A}
\label{eq:4.2}
\end{equation}
As it turns out, the super Ashtekar connection defines a so-called \emph{generalized super Cartan connection}, so that, via the correspondence Cartan $\leftrightarrow$ Ehresmann, it gives rise to a proper super connection 1-form on the associated $\mathrm{OSp}(1|2)_{\mathbb{C}}$-bundle \cite{Eder:2021ans,Eder:2020erq}. By applying the projection \eqref{eq:4.1} on the Cartan curvature \eqref{eq:3.6}, we then find for the Lorentzian sub components  
\begin{equation}
(\mathbf{P}^+F(\mathcal{A}))^i=F(A^+)^i+\frac{i}{L}\psi^A\wedge\psi^B\tau^i_{AB}+\frac{1}{2L^2}\Sigma^i=F(\mathcal{A}^+)^i+\frac{1}{2L^2}\Sigma^i
\label{eq:4.3}
\end{equation}
with $F(\mathcal{A}^+)$ the associated curvature of $\mathcal{A}^+$. Here, $\Sigma^{AB}=\Sigma^i\tau_i^{AB}$ denotes the self-dual part of $\Sigma^{AA'BB'}:=e^{AA'}\wedge e^{BB'}$ which, due to antisymmetry, can be decomposed according to  
\begin{equation}
\Sigma^{AA'BB'}=\epsilon^{AB}\Sigma^{A'B'}+\epsilon^{A'B'}\Sigma^{AB}
\label{eq:4.4}
\end{equation}
such that $\Sigma^{AB}:=\frac{1}{2}\epsilon_{A'B'}\Sigma^{AA'BB'}$. Moreover, for the chiral odd components, we find 
\begin{equation}
(\mathbf{P}^+F(\mathcal{A}))^A=D^{(A^+)}\psi^A+\frac{1}{2L}\chi^A=F(\mathcal{A}^+)^A+\frac{1}{2L}\chi^A
\label{eq:4.5}
\end{equation}
where we set $\chi:=-\boldsymbol{\gamma}\wedge\psi$ such that $\chi_A=e_{AA'}\wedge\bar{\psi}^{A'}$. Thus, defining $\mathcal{E}:=\Sigma^i T_i^++L\chi^AQ_A$ which will also be called the \emph{super electric field}, this yields   
%\begin{equation}
%F(\mathcal{A}^+)^{AB}=F(A^+)^{AB}+\frac{i}{2L}\psi^A\wedge\psi^B
%\label{eq:}
%\end{equation}
\begin{equation}
\mathbf{P}^+F(\mathcal{A})=F(\mathcal{A}^+)+\frac{1}{2L^2}\mathcal{E}
\label{eq:4.6}
\end{equation} 
Inserting this expression into the Holst-MacDowell-Mansouri action \eqref{eq:3.5} for $\beta=-i$, this gives
\begin{align}
\braket{F(\mathcal{A})\wedge F(\mathcal{A})}_{\beta}=&i\braket{(F(\mathcal{A}^+)+\frac{1}{2L^2}\mathcal{E})\wedge(F(\mathcal{A}^+)+\frac{1}{2L^2}\mathcal{E})}\nonumber\\
=&\frac{i}{L^2}\braket{\mathcal{E}\wedge F(\mathcal{A}^+)}+\frac{i}{4L^4}\braket{\mathcal{E}\wedge\mathcal{E}}+i\braket{F(\mathcal{A}^+)\wedge F(\mathcal{A}^+)}
\end{align}
such that
\begin{align}
S(\mathcal{A})=\frac{i}{\kappa}\int_{M}{\braket{\mathcal{E}\wedge F(\mathcal{A}^+)}+\frac{1}{4L^2}\braket{\mathcal{E}\wedge\mathcal{E}}}+S_{\text{bdy}}
\label{eq:4.6.1}
\end{align}
with a boundary term $S_{\text{bdy}}$ taking the form
\begin{equation}
S_{\text{bdy}}(\mathcal{A}^+)=\frac{iL^2}{\kappa}\int_{M}{\braket{F(\mathcal{A}^+)\wedge F(\mathcal{A}^+)}}=\frac{iL^2}{\kappa}\int_{\partial M}{\braket{\mathcal{A}^+\wedge\mathrm{d}\mathcal{A}^++\frac{1}{3}\mathcal{A}^+\wedge[\mathcal{A}^+\wedge\mathcal{A}^+]}}
\label{eq:4.7}
\end{equation}
where we used \eqref{eq:A6} which now holds due to $\mathrm{OSp}(1|2)_{\mathbb{C}}$-invariance. Thus, as we see, in the chiral theory, the Holst-MacDowell-Mansouri action becomes manifest $\mathrm{OSp}(1|2)_{\mathbb{C}}$ gauge invariant and the boundary term takes the form of a $\mathrm{OSp}(1|2)_{\mathbb{C}}$ super Chern-Simons action with (complex) Chern-Simons level $k=i4\pi L^2/\kappa=-i12\pi/\kappa\Lambda$ with $\Lambda$ the cosmological constant.\\
Finally, for the last part of this section, we want to explicitly show that the full action \eqref{eq:4.6.1} is indeed invariant under local supersymmetry transformations. Therefore, let us further evaluate the bulk term in \eqref{eq:4.6.1}. Using \eqref{eq:4.3} and \eqref{eq:4.5}, we find   
\begin{align}
\braket{\mathcal{E}\wedge F(\mathcal{A}^+)}&=\Sigma^{AB}\wedge F(A^+)_{AB}+\frac{i}{2L}\Sigma^{AB}\wedge\psi_A\wedge\psi_B+i\chi_A\wedge D^{(A^+)}\psi^A,
\end{align}
as well as
\begin{equation}
\braket{\mathcal{E}\wedge\mathcal{E}}=\Sigma^{AB}\wedge\Sigma_{AB}+iL\chi_A\wedge\chi^A,
\label{eq:4.8}
\end{equation}
so that the bulk action can be written in the form
\begin{align}
S(\mathcal{A})=&\frac{i}{\kappa}\int_{M}{\Sigma^{AB}\wedge F(A^+)_{AB}+i\chi_A\wedge D^{(A^+)}\psi^A}\nonumber\\
&+\frac{i}{2L}\Sigma^{AB}\wedge\psi_A\wedge\psi_B+\frac{i}{4L}\chi_A\wedge\chi^A+\frac{1}{4L^2}\Sigma^{AB}\wedge\Sigma_{AB}.
\end{align}
This is precisely the form of the action of chiral $\mathcal{N}=1$, $D=4$ AdS supergravity as stated e.g. in \cite{Sano:1992jw,Fulop:1993wi}. 

In the Weyl-representation of the gamma matrices, the Majorana spinor $\epsilon$ generating supersymmetry transformations splits into a left- and right-handed Weyl spinor $\epsilon=(\epsilon^A,\epsilon_{A'})^T$. We will say that transformations associated with the former are left-handed supersymmetry transformations, whereas the latter will be called right-handed supersymmetry transformations. 
According to the general discussion in section \ref{N=1 SUGRA with inner bdy}, it follows that under left-supersymmetry transformations corresponding to $\epsilon=(\epsilon^A,0)^T$, super Ashtekar connection and electric field transform via
\begin{equation}
    \delta_{\epsilon}\mathcal{A^+}=D^{(\mathcal{A}^+)}\epsilon\quad\text{ and }\quad \delta_{\epsilon}\mathcal{E}=-[\epsilon,\mathcal{E}]
\end{equation}
respectively, and therefore correspond to ordinary $\mathrm{OSp}(1|2)_{\mathbb{C}}$-gauge transformations under which the action \eqref{eq:4.6.1} is manifestly invariant. Note that this is true even off-shell, i.e. without $\omega$ satisfying its field equation. Thus, in the chiral theory, it follows that a sub part of the full SUSY transformations becomes a true gauge symmetry of the theory!\\
It remains to show that the  action is invariant under right-SUSY transformations corresponding to some (anticommutative) right-handed Weyl spinor $\epsilon=(0,\epsilon_{A'})^T$. In that case, from \eqref{eq:1.3} we deduce 
\begin{equation}
\delta_{\epsilon}e^{AA'}=i\psi^A\epsilon^{A'},\quad\delta_{\epsilon}\psi^A=-\frac{1}{2L}e^{AA'}\epsilon_{A'},\quad\delta_{\epsilon}\psi_{A'}=D^{(A^-)}\epsilon_{A'},\quad\delta_{\epsilon}A^{+}=0.
\label{eq:4.9}
\end{equation}
Using \eqref{eq:4.9}, it follows that the variation of the super electric field takes the form
\begin{equation}
\delta_{\epsilon}\chi_A=i\psi_A\epsilon_{A'}\wedge\bar{\psi}^{A'}+e_{AA'}\wedge D^{(A^-)}\epsilon_{A'}=D^{(A^+)}\eta_A+(D^{(\omega)}e_{AA'}+i\psi_A\wedge\bar{\psi}_{A'})
\label{eq:4.10}
\end{equation}
where $\eta^A:=e^{AA'}\epsilon_{A'}$, as well as
\begin{equation}
\delta_{\epsilon}\Sigma^{AB}=i\psi^{(A}\wedge\eta^{B)}
\label{eq:4.11}
\end{equation}
In what follows, we want to assume that the self-dual Ashtekar connection $A^+$ satisfies its field equation. Therefore, it is important to note that the field equations of both $A^+$ and $\psi^A$ are altered due to the appearence of additional boundary terms in the full action. More precisely, if one varies \eqref{eq:4.6.1} with respect to super Ashtekar connection $\mathcal{A}^+$, one finds that
\begin{align}
    \delta S(\mathcal{A})&=\frac{i}{\kappa}\int_M\braket{D^{(\mathcal{A}^+)}\delta\mathcal{A}^+\wedge\mathcal{E}}+\frac{2iL^2}{\kappa}\int_M\braket{D^{(\mathcal{A}^+)}\delta\mathcal{A}^+\wedge F(\mathcal{A}^+)}\nonumber\\
    &=\frac{i}{\kappa}\int_M\braket{\delta\mathcal{A}^+\wedge D^{(\mathcal{A}^+)}\mathcal{E}}+\frac{i}{\kappa}\int_{\partial M}\braket{\delta\mathcal{A}^+\wedge[\mathcal{E}+2L^2F(\mathcal{A}^+)]}=0
    \label{eq:4.11.1}
\end{align}
where we have integrated by parts and used the Bianchi identity $D^{(\mathcal{A}^+)}F(\mathcal{A}^+)=0$. Hence, the EOM of $\mathcal{A}^+$ are unaltered provided that the boundary contribution in \eqref{eq:4.11.1} vanishes, i.e.
\begin{equation}
    \underset{\raisebox{1pt}{$\Longleftarrow$}}{F(\mathcal{A}^{+})}=-\frac{1}{2L^2}\underset{\raisebox{1pt}{$\Leftarrow$}}{\mathcal{E}}
    \label{eq:4.11.2}
\end{equation}
where the arrow denotes the pullback to the boundary. In the following, let us assume that the boundary condition \eqref{eq:4.11.2} is satisfied. Then, modulo the field equations of $A^+$ given by $D^{(A^+)}\Sigma^{AB}+i\psi^{(A}\wedge\chi^{B)}=0$ and which, as discussed in detail in \cite{Jacobson:1987cj,Eder:2020erq}, turn out to be equivalent to the EOM of $\omega$, i.e.
\begin{equation}
D^{(\omega)}e_{AA'}+i\psi_A\wedge\bar{\psi}_{A'}=0
\end{equation}
it follows that
\begin{align}
\delta_{\epsilon}(\Sigma^{AB}\wedge F(A^+)_{AB})=i\psi^{(A}\wedge\eta^{B)}\wedge F(A^+)_{AB}%+\Sigma^{AB}\wedge D^{(A^+)}\delta_{\epsilon}A^+_{AB}\nonumber\\
%&=i\psi^{(A}\wedge\eta^{B)}\wedge F(A^+)_{AB}+\mathrm{d}(\Sigma^{AB}\wedge\delta_{\epsilon}A^+_{AB})-D^{(A^+)}\Sigma^{AB}\wedge\delta_{\epsilon}A^+_{AB}
\label{eq:4.12}
\end{align}
On the other hand, we have
\begin{align}
\delta_{\epsilon}(\chi_A\wedge D^{(A^+)}\psi^A)=&D^{(A^+)}\eta_A\wedge D^{(A^+)}\psi^A-\frac{1}{2L}\chi_A\wedge D^{(A^+)}\eta^A\nonumber\\%+\chi^A\wedge\tensor{{\delta_{\epsilon}A^+}}{^A_B}\wedge\psi^B\nonumber\\
=&\mathrm{d}(\eta_A\wedge D^{(A^+)}\psi^A)+\eta_A\wedge D^{(A^+)}D^{(A^+)}\psi^A-\frac{1}{2L}\chi_A\wedge D^{(A^+)}\eta^A\nonumber\\
%&-\psi^{(A}\wedge\chi^{B)}\wedge\delta_{\epsilon}A^+_{AB}\nonumber\\
=&\mathrm{d}(\eta_A\wedge D^{(A^+)}\psi^A)-\psi_{(A}\wedge\eta_{B)}\wedge F(A^+)^{AB}-\frac{1}{2L}\chi_A\wedge D^{(A^+)}\eta^A%\nonumber\\
%&-\psi^{(A}\wedge\chi^{B)}\wedge\delta_{\epsilon}A^+_{AB}
\label{eq:4.13}
\end{align}
as well as 
\begin{align}
\delta_{\epsilon}(\Sigma^{AB}\wedge\psi_A\wedge\psi_B)=-\frac{1}{L}\Sigma^{AB}\wedge\eta_{(A}\wedge\psi_{B)}
\label{eq:4.14}
\end{align}
Finally, using $\delta_{\epsilon}(\chi_A\wedge\chi^A)=2\chi_A\wedge D^{(A^+)}\eta^{A}$ and $\delta_{\epsilon}(\Sigma^{AB}\wedge\Sigma_{AB})=2i\Sigma^{AB}\wedge\eta_{(A}\wedge\psi_{B)}$, we finally obtain for the variation of the full Lagrangian under right-handed SUSY transformations
\begin{align}
\delta_{\epsilon}\mathscr{L}=i\mathrm{d}(\eta_A\wedge D^{(A^+)}\psi^A)+\delta_{\epsilon}\mathscr{L}_{\text{bdy}}
\label{eq:4.15}
\end{align} 
where, by the Bianchi identity, the variation of the boundary term can be written in the form
\begin{align}
\delta_{\epsilon}\mathscr{L}_{\text{bdy}}&=L^2\delta_{\epsilon}\braket{F(\mathcal{A}^+)\wedge F(\mathcal{A}^+)}=2L^2\braket{D^{(\mathcal{A}^+)}\delta_{\epsilon}\mathcal{A}^+\wedge F(\mathcal{A}^+)}\nonumber\\
&=2L^2\mathrm{d}\!\braket{\delta_{\epsilon}\mathcal{A}^+\wedge F(\mathcal{A}^+)}+2L^2\braket{\delta_{\epsilon}\mathcal{A}^+\wedge D^{(\mathcal{A}^+)}F(\mathcal{A}^+)}\nonumber\\
&=2L^2\mathrm{d}\!\braket{\delta_{\epsilon}\psi\wedge D^{(A^+)}\psi}=-L\mathrm{d}\!\braket{\eta\wedge D^{(A^+)}\psi}=-i\mathrm{d}(\eta_A\wedge D^{(A^+)}\psi^A).
\label{eq:4.16}
\end{align} 
Thus, combining \eqref{eq:4.16} and \eqref{eq:4.15}, we see that the variation of the boundary term cancels exactly with the respective contribution of the bulk Lagrangian, finally yielding 
\begin{align}
\delta_{\epsilon}\mathscr{L}=0.
\label{eq:4.17}
\end{align} 
This proves that, provided boundary condition \eqref{eq:4.11.2} is satisfied, the full action is indeed invariant under local SUSY transformations at the boundary. Moreover, from the previous computations, we infer that, in the presence of boundaries, the boundary contributions \eqref{eq:4.7} taking the form of a $\mathrm{OSp}(1|2)_{\mathbb{C}}$ super Chern-Simons action to the full action \eqref{eq:4.6.1} are in fact unique if one requires both manifestly $\mathrm{OSp}(1|2)_{\mathbb{C}}$-gauge invariance and invariance under local right-handed SUSY transformations at the boundary. As we will see in the next sections, these observations even carry over to supergravity theories with extended supersymmetry.
\begin{remark}
As an aside, note that one can introduce a new independent 2-form field $\mathcal{B}$, also simply called the \emph{B-field}, satisfying the simplicity constraint $\mathcal{B}:=\mathcal{E}$. In this way, it then follows that one can rewrite the bulk term in \eqref{eq:4.6.1} in terms of a constrained \emph{super BF-action} with a non-vanishing cosmological constant \cite{Ezawa:1995nj,Ling:2000ss,Ling:2003yw}.
\end{remark}
As a last step, let us briefly comment on the canonical analysis of the theory as well as the boundary conditions which couple the bulk and boundary degrees of freedom. Therefore, we split the full action \eqref{eq:4.6.1} into a bulk and boundary term such that $S_{\text{bulk}}+S_{\text{bdy}}$ with $S_{\text{bdy}}$ given by \eqref{eq:4.7}. Similar as above, the variation of the bulk contribution with respect to $\mathcal{A}^+$ then yields
\begin{equation}
    \delta S_{\text{bulk}}=\frac{i}{\kappa}\int_M\braket{D^{(\mathcal{A}^+)}\delta\mathcal{A}^+\wedge\mathcal{E}}=:\mathrm{d}\Theta+\frac{i}{\kappa}\int_M\braket{\delta\mathcal{A}^+\wedge D^{(\mathcal{A}^+)}\mathcal{E}}
    \label{eq:4.18}
\end{equation}
Here, $\Theta(\delta)$ denotes the pre-symplectic potential inducing the bulk pre-symplectic structure $\Omega_{\text{bulk}}=\mathbbm{d}\Theta$
\begin{align}
    \Omega_{\text{bulk}}(\delta_1,\delta_2)&=\frac{2i}{\kappa}\int_{\Sigma}\braket{\delta_{[1}\mathcal{A}^+\wedge\delta_{2]}\mathcal{E}}
    \label{eq:4.19}
\end{align}
and, as a consequence, $(\mathcal{A}^+,\mathcal{E})$ (or rather their pullback to $\Sigma$) define canonically conjugate variables of the canonical phase space. Moreover, from \eqref{eq:4.18}, we can immediately read off the Gauss constraint which takes the form
\begin{equation}
    \mathcal{G}(\Lambda)=\int_{\Sigma}\braket{D^{(\mathcal{A}^+)}\mathcal{E},\Lambda}
\end{equation}
where we have chosen an arbitrary $\mathfrak{osp}(1|2)_{\mathbb{C}}$-valued smearing function $\Lambda$. As can be checked by direct computation, the Gauss constraint satisfies the constraint algebra $\{\mathcal{G}(\Lambda),\mathcal{G}(\Lambda')\}=\mathcal{G}([\Lambda,\Lambda'])$ and therefore generates local $\mathrm{OSp}(1|2)_{\mathbb{C}}$-gauge transformations.\\
On the other hand, as already explained above, the boundary contribution to \eqref{eq:4.6.1} is given by the action corresponding to a $\mathrm{OSp}(1|2)_{\mathbb{C}}$ super Chern-Simons theory. As a result, the pre-symplectic structure of the full theory takes the form 
\begin{align}
    \Omega(\delta_1,\delta_2)&=\frac{2i}{\kappa}\int_{\Sigma}\braket{\delta_{[1}\mathcal{A}^+\wedge\delta_{2]}\mathcal{E}}-\frac{k}{2\pi}\int_{\partial\Sigma}\braket{\delta_{[1}\mathcal{A}^+\wedge\delta_{2]}\mathcal{A}^+}
    \label{eq:4.19.1}
\end{align}
Furthermore, the decomposition of the full action into bulk and boundary terms leads to a matching condition on the boundary between bulk and boundary degrees of freedom. This is equivalent to requiring consistency with the equation of motion of the full theory, i.e. $\delta S=\delta S_{\text{bulk}}+\delta S_{\text{bdy}}=0$ which leads back to boundary condition
 \eqref{eq:4.11.2}. As we see, this condition arises quite naturally from the requirement of supersymmetry invariance at the boundary and in fact, based on the previous observations, even turns out to be unique.
\begin{remark}
Note that one can also rewrite the boundary condition \eqref{eq:4.11.2} in the equivalent form
\begin{equation}
    \underset{\raisebox{1pt}{$\Longleftarrow$}}{F(\mathcal{A}^{+})}+\frac{1}{2L^2}\underset{\raisebox{1pt}{$\Leftarrow$}}{\mathcal{E}}=0\quad\Leftrightarrow\quad\underset{\raisebox{1pt}{$\Longleftarrow$}}{\mathbf{P}^+F(\mathcal{A})}=0
    \label{eq:4.21}
\end{equation}
according to the identity \eqref{eq:4.6}. Furthermore, taking the complex conjugate of \eqref{eq:4.21} yields $\mathbf{P}^-F(\mathcal{A})=0$. Hence, when combined together, this in turn gives
\begin{equation}
    \underset{\raisebox{1pt}{$\Longleftarrow$}}{\mathbf{P}^+F(\mathcal{A})}+\underset{\raisebox{1pt}{$\Longleftarrow$}}{\mathbf{P}^-F(\mathcal{A})}=0\quad\Leftrightarrow\quad \underset{\raisebox{1pt}{$\Longleftarrow$}}{F(\mathcal{A})}^{IJ}=0\text{ and }\underset{\raisebox{1pt}{$\Longleftarrow$}}{F(\mathcal{A})}^{\alpha}=0
    \label{eq:4.211}
\end{equation}
that is, the Cartan curvature associated to the (full) super Cartan connection $\mathcal{A}$ is constrained to vanish at the boundary. This is precisely the boundary condition as derived in \cite{Andrianopoli:2014aqa} in context of the non-chiral theory.  
\end{remark}

\begin{remark}
The derivation of the Holst-MacDowell-Mansouri action via the $\beta$-deformed inner product as described in this article also gives an elegant approach to the 'double chiral' action as considered e.g. in \cite{Ling:2000ss}. One notices that the standard action of $\mathcal{N}=1$ SUGRA (modulo boundary terms) arises from \eqref{eq:3.5} in the limit $\beta\rightarrow\infty$. On the other hand, one has $\mathbf{P}_{i}+\mathbf{P}_{-i}=2\mathbf{P}_{\infty}$ where $\mathbf{P}_{\pm i}=\mp i\mathbf{P}^{\mp}$ with $\mathbf{P}^{\mp}$ defining projections from $\mathfrak{osp}(1|4)$ onto two chiral copies of $\mathfrak{osp}(1|2)_{\mathbb{C}}$. The $\mathcal{N}=1$ SUGRA action then decomposes as
\begin{equation}
    2S_{MM}(\mathcal{A})=S_{H-MM}^{\beta=i}(\mathcal{A})+S_{H-MM}^{\beta=-i}(\mathcal{A})
\end{equation}
and thus splits into two chiral acions of the form \eqref{eq:4.6.1}. These actions can be expressed in terms of the graded Ashtekar connections $\mathcal{A}^{-}$ and $\mathcal{A}^{+}$, respectively.
\end{remark}

\section{$\mathcal{N}=2$ pure SUGRA with boundaries}\label{Section:Pure N=2}
Similar to the non-extended case, pure $\mathcal{N}=2$, $D=4$ anti-de Sitter supergravity can be described as a super Cartan geometry modeled over the super Klein geometry\footnote{Note that, in the ungauged theory, the full $R$-symmetry group is given by the unitary group $\mathrm{U}(2)$. However, in case of AdS supergravity, due to the appearence of the so-called \emph{Fayet-Iliopoulos (FI) term}, it follows that this group is broken yielding an effective $\mathrm{SO}(2)\cong\mathrm{U}(1)$ gauge symmetry of the theory (see e.g. \cite{Andrianopoli:1996cm,Andrianopoli:2014aqa,Andrianopoli:2020zbl} for more details).} $(\mathrm{OSp}(2|4),\mathrm{SO}(2)\times\mathrm{Spin}^+(1,3))$ corresponding to extended super anti-de Sitter space. In this case, since $\mathfrak{osp}(2|4)\cong\mathbb{R}^{1,3}\oplus\mathfrak{spin}^+(1,3)\oplus\Delta_{\mathbb{R}}^2\oplus\mathfrak{u}(1)$, the super Cartan connection $\mathcal{A}$ takes the form
\begin{equation}
\mathcal{A}=e^IP_I+\frac{1}{2}\omega^{IJ}M_{IJ}+\hat{A}T+\Psi_r^{\alpha}Q_{\alpha}^r
\label{eq:5.1}
\end{equation}   
In particular, besides the spin connection $\omega$, the super Cartan connection contains an additional $\mathrm{U}(1)$ gauge field $\hat{A}\equiv\hat{A}T$ also referred to as the \emph{graviphoton field} with $T:=T^{12}=-T^{21}$. Moreover, the supermultiplet consists of two Majorana gravitinos which we denote by capital letters $\Psi_r$, $r=1,2$ to simplify notation. In this form, the $R$-symmetry index is raised and lowered with the Kronecker symbol $\delta_{rs}$. On the other hand, we denote the individual chiral components of the Majorana fermions by lower case letters $\psi^r$ and $\psi_r$, respectively, where the position of the $R$-symmetry index now explicitly indicates the chirality:
\begin{equation}
\psi^r:=\frac{\mathds{1}+\gamma_*}{2}\Psi_r,\text{ and }\psi_r:=\frac{\mathds{1}-\gamma_*}{2}\Psi_r
\label{eq:5.2}
\end{equation}  
for $r=1,2$ denote the left-handed and right-handed components of the Majorana fermions, respectively. Similar to the $\mathcal{N}=1$ case, the horizontal 1-forms combine to the super soldering form $E:=e^IP_I+\Psi_r^{\alpha}Q_{\alpha}^r$ which provides a local identification of the underlying (curved) supermanifold $\mathcal{M}$ with the flat model given by the extended super anti-de Sitter space. In particular, it induces an isomorphism
\begin{equation}
E:\,\Gamma(T\mathcal{M})\stackrel{\sim}{\rightarrow}\Gamma(\mathrm{Ad}(\mathcal{P})),\,X\mapsto\braket{X|E}
\label{eq:5.3}
\end{equation}
between smooth vector fields on $\mathcal{M}$ and sections of the \emph{Adjoint-bundle} $\mathcal{P}\times_{\mathrm{Ad}}\mathfrak{osp}(2|4)$, where $\mathcal{P}$ is the underlying $\mathrm{U}(1)\times\mathrm{Spin}^+(1,3)$ principal super fiber bundle over which the super Cartan geometry is defined. By the \emph{rheonomy principle}, the fields are uniquely fixed by their pullback to the underlying bosonic spacetime manifold $M$. Hence, choosing a local section $s:\,M\rightarrow P\subset\mathcal{P}$ of the bosonic subbundle $P$, the action of the theory takes the form
\begin{equation}
S(\mathcal{A})=\frac{L^2}{\kappa}\int_{M}{s^*\mathscr{L}}
\label{eq:5.4}
\end{equation}
where the Lagrangian $\mathscr{L}$ is a horizontal form living on the bundle which, when adapted to our choice of conventions and pulled back to $M$, takes the form\footnote{Due to the appearance of the Hodge-star operator in the Maxwell-kinetic term in the Lagrangian \eqref{eq:5.5}, in this form, the Lagrangian can only be defined on the underlying spacetime manifold. This is related to the lack of top-degree forms on supermanifolds (for an alternative approach towards top-degree forms on supermanifolds using the concept of \emph{integral forms} see e.g. \cite{Castellani:2014goa,Catenacci:2018xsv}). In order to extend \eqref{eq:5.5} to the whole supermanifold, one works in the so-called first-order formalism in the $\mathrm{U}(1)$-sector by introducing additional fields (auxiliary fields). By solving the equations of motion of these additional fields, one regains the original action \eqref{eq:5.4} (see e.g. \cite{Andrianopoli:2014aqa,Andrianopoli:2020zbl}).} \cite{Andrianopoli:1996cm,Andrianopoli:2014aqa,Andrianopoli:2020zbl}
\begin{align}
s^*\mathscr{L}=&\frac{1}{4L^2}\Sigma^{IJ}\wedge F(\omega)^{KL}\epsilon_{IJKL}-\frac{i}{2L^2}\bar{\Psi}^r\wedge\boldsymbol{\gamma}\gamma_{*}\wedge\nabla\Psi_r-\frac{1}{8L^3}\bar{\Psi}^r\wedge\gamma^{KL}\Psi_r\wedge\Sigma^{IJ}\epsilon_{IJKL}\nonumber\\
&+\frac{1}{8L^4}\Sigma^{IJ}\wedge\Sigma^{KL}\epsilon_{IJKL}+\frac{i}{4L^2}\left(\mathrm{d}\hat{A}+\frac{1}{4}\bar{\Psi}^r\wedge\Psi^s\epsilon_{rs}\right)\wedge\bar{\Psi}^p\wedge\gamma_{*}\Psi^q\epsilon_{pq}-\frac{1}{4L^2}\hat{F}\wedge\star \hat{F}
\label{eq:5.5}
\end{align}
where, again, $\Sigma^{IJ}=e^I\wedge e^J$ whereas $\nabla\Psi_r$ and $\hat{F}$ (resp. $\star \hat{F}$) are defined via equation \eqref{eq:6.7} and \eqref{eq:6.8} in section \ref{Holst N=2} below. In contrast to the $\mathcal{N}=1$ case, supersymmetry transformations no longer have the simple interpretation in terms of gauge transformations on the associated $\mathrm{OSp}(2|4)$-bundle. Instead, according to the D'Auria-Fré approach, one regards them as certain superdiffeomorphisms along the odd directions of the supermanifold. More precisely, SUSY transformations correspond to smooth vector fields $\epsilon\in\Gamma(T\mathcal{M})$ such that $i_{\epsilon}e^I=0$ and
\begin{equation}
s^*(\delta_{\epsilon}\mathscr{L}):=s^*(L_{\epsilon}\mathscr{L})=s^*(i_{\epsilon}\mathrm{d}\mathscr{L}+\mathrm{d}i_{\epsilon}\mathscr{L})=0
\label{eq:5.6}
\end{equation}
for any section $s:\,M\rightarrow P\subset\mathcal{P}$. The super Cartan connection transforms as
\begin{equation}
\delta_{\epsilon}\mathcal{A}=i_{\epsilon}F(\mathcal{A})+D^{(\mathcal{A})}(i_{\epsilon}E)
\label{eq:5.7}
\end{equation}
Since $\epsilon$ is horizontal, the curvature contributions in \eqref{eq:5.7}, in general, no longer vanishes in contrast to pure gauge transformations. Moreover, in order for $\epsilon$ to describe a symmetry of the theory, this imposes constraints on the curvature, the so-called \emph{rheonomy conditions}. Instead of deriving the explicit form of the SUSY transformations of the theory in what follows (see for instance \cite{Andrianopoli:1996cm,Andrianopoli:2014aqa,Castellani:1991et,Andrianopoli:2020zbl} for more details), let us finally comment on the possible boundary terms to be added to \eqref{eq:5.5} such that local supersymmetry is preserved.\\
In fact, as shown in \cite{Andrianopoli:2014aqa}, the most general form of the boundary terms which are compatible with the symmetries of the bulk action \eqref{eq:5.5} and which, in particular, are compatible with local supersymmetry at the boundary take the form\footnote{To be more precise, it follows that the most general ansatz has to be of the form
\begin{equation*}
    \mathscr{L}_{\text{bdy}}=C_1F(\omega)^{IJ}\wedge F(\omega)^{KL}\epsilon_{IJKL}+C_2\mathrm{d}(\bar{\Psi}^r\wedge\gamma_{*}\nabla\Psi_r)+C_3\mathrm{d}(\hat{A}\wedge\mathrm{d}\hat{A})
\end{equation*}
for any constant coefficients $C_1,C_2$ and $C_3$. Requiring invariance of the full action under local supersymmetry transformations then fixes the constants to the particular values $C_1=\frac{1}{8}$, $C_2=\frac{i}{2L}$ and $C_3=0$ leading to \eqref{eq:5.8}.}
\begin{align}
\mathscr{L}_{\text{bdy}}=&\frac{1}{8}F(\omega)^{IJ}\wedge F(\omega)^{KL}\epsilon_{IJKL}+\frac{i}{2L}\nabla\bar{\Psi}^r\wedge\gamma_{*}\nabla\Psi_r-\frac{i}{8L}F(\omega)^{IJ}\wedge\bar{\Psi}^r\wedge\gamma_{*}\gamma_{IJ}\Psi_r\nonumber\\
&-\frac{i}{4L^2}\mathrm{d}\hat{A}\wedge\bar{\Psi}^p\wedge\gamma_{*}\Psi^q\epsilon_{pq}
\label{eq:5.8}
\end{align}
When added to \eqref{eq:5.5}, one then recognizes that the resulting action has a very intriguing structure similar to the MacDowell-Mansouri action of $\mathcal{N}=1$ AdS supergravity as discussed in the previous sections \cite{Andrianopoli:2014aqa,Andrianopoli:2020zbl}.

\subsection{Holst action for $\mathcal{N}=2$ pure SUGRA}\label{Holst N=2}
We want to derive a Holst variant of the action \eqref{eq:5.5} corresponding to $\mathcal{N}=2$, $D=4$ AdS supergravity for arbitrary Barbero-Immirzi parameters $\beta$ including the boundary terms \eqref{eq:5.8}. We therefore follow the ideas in section \ref{Holst N=1} and introduce a $\beta$-deformed inner product. To this end, according to the decomposition \eqref{eq:5.1} of the super Cartan connection, let us define an operator $\mathbf{P}_{\beta}:\,\Omega^2(M,\mathfrak{g})\rightarrow\Omega^2(M,\mathfrak{g})$ on the space of differential 2-forms with values in the super Lie algebra $\mathfrak{g}:=\mathfrak{osp}(2|4)$ as follows
\begin{equation}
\mathbf{P}_{\beta}:=\underline{\mathbf{0}}\oplus\mathcal{P}_{\beta}\oplus\mathbb{P}_{\beta}\oplus\mathcal{P}_{\beta}\oplus\mathcal{P}_{\beta},\text{ where }\quad\mathbb{P}_{\beta}:=\frac{1}{2\beta}\left(1+\beta\star\right)
\label{eq:6.1}
\end{equation}
with $\star:\,\Omega^p(M)\rightarrow\Omega^{4-p}(M)$, for $0\leq p\leq 4$ denoting the \emph{Hodge star} operator on the bosonic spacetime manfifold $M$ (trivially extended to $\mathfrak{g}$-valued, in fact even Grassmann-valued, differential forms) which, in case of Lorentzian signature and even spacetime dimensions, satisfies 
\begin{equation}
\star^2|_{\Omega^p(M)}=(-1)^{p+1},\,\forall 0\leq p\leq 4 
\label{eq:6.2}
\end{equation}
Similar to the general discussion in the appendix \ref{Chern-Simons}, the operator \eqref{eq:6.1} can be used to introduce an inner product on $\Omega^2(M,\mathfrak{g})$ setting 
\begin{align}
\braket{\cdot\wedge\cdot}_{\beta}:\,\Omega^2(M,\mathfrak{g})\times\Omega^2(M,\mathfrak{g})&\rightarrow\Omega^4(M)\nonumber\\
(\omega,\eta)&\mapsto\mathrm{str}(\omega\wedge\mathbf{P}_{\beta}\eta)
\label{eq:6.3}
\end{align}
Using this inner product, we define the Holst-MacDowell-Mansouri action of $\mathcal{N}=2$, $D=4$ AdS supergravity action as follows
\begin{equation}
S(\mathcal{A})=\frac{L^2}{\kappa}\int_{M}{\braket{F(\mathcal{A})\wedge F(\mathcal{A})}_{\beta}}
\label{eq:6.4}
\end{equation}
with $F(\mathcal{A})$ the associated Cartan curvature. Using the commutation relations \eqref{eq:C8a}-\eqref{eq:C8d} for $\mathcal{N}=2$, it follows that the translational components of the curvature take the form
\begin{align}
F(\mathcal{A})^I&=\mathrm{d}e^I+\tensor{\omega}{^I_J}\wedge e^J+\frac{1}{4}((-1)^{|Q_{\alpha}||Q_{\beta}|}\Psi^{\alpha}_r\wedge\Psi^{\beta}_s\otimes[Q^r_{\alpha},Q^s_{\beta}])^{I}\nonumber\\
&=\Theta^{(\omega) I}-\frac{1}{4}\bar{\Psi}^r\wedge\gamma^I\Psi_r
\label{eq:6.5}
\end{align}
since $(-1)^{|Q_{\alpha}||Q_{\beta}|}=-1$, with $\Theta^{(\omega)}$ the torsion 2-form associated to the spin connection $\omega$. For the Lorentzian components, we find
\begin{align}
F(\mathcal{A})^{IJ}&=\mathrm{d}\omega^{IJ}+\tensor{\omega}{^I_K}\wedge\omega^{KJ}+\frac{1}{2L^2}e^I\wedge e^J-\frac{1}{2}(\Psi_r^{\alpha}\wedge\Psi_s^{\beta}\otimes[Q^r_{\alpha},Q^s_{\beta}])^{IJ}\nonumber\\
&=F(\omega)^{IJ}+\frac{1}{L^2}\Sigma^{IJ}-\frac{1}{4L}\bar{\Psi}^r\wedge\gamma^{IJ}\Psi_r
\label{eq:6.6}
\end{align}
Moreover, for the odd part, we obtain, using $\hat{A}:=\frac{1}{2}\hat{A}^{rs}T_{rs}$ for the $\mathrm{U}(1)$ gauge field,
\begin{align}
F(\mathcal{A})^{\alpha}_r&=D^{(\omega)}\Psi_r^{\alpha}+\frac{1}{2L}\hat{A}\epsilon_{rs}\wedge\Psi^{\alpha s}-\frac{1}{2L}e^I\wedge\tensor{(\gamma_{I})}{^{\alpha}_{\beta}}\Psi_r^{\beta}=:\nabla\Psi_r^{\alpha}-\frac{1}{2L}e^I\wedge\tensor{(\gamma_{I})}{^{\alpha}_{\beta}}\Psi_r^{\beta}
\label{eq:6.7}
\end{align}
Finally, for the $\mathrm{U}(1)$ components, we get
\begin{align}
\hat{F}:=\frac{1}{2}F(\mathcal{A})^{rs}\epsilon_{rs}=\mathrm{d}\hat{A}+\frac{1}{2}\bar{\Psi}^r\wedge\Psi^s\epsilon_{rs}
\label{eq:6.8}
\end{align}
We need to show that the field equations arsing from the deformed action \eqref{eq:6.4} are indeed independent of the choice of the Barabero-Immirzi parameter and thus, in particular, coincide with the field equations of the standard SUGRA theory. We therefore prove that the action at second order, i.e. provided $\omega$ satisfies its field equations, reduces to the action \eqref{eq:5.5} together with the boundary term \eqref{eq:5.8}. This is equivalent to requiring that the supertorsion \label{eq:6.5} of $\mathcal{A}$ vanishes, i.e. $F(\mathcal{A})^I=0$, and, when reinserting back into \eqref{eq:6.4}, all $\beta$-dependent terms become purely topological.\\
Therefore, let us expand the action \eqref{eq:6.4}. Using the curvature expressions \eqref{eq:6.6}-\eqref{eq:6.8} as well as $\braket{T,T}_{\beta}=-\frac{1}{2L^2}$, which can be checked by direct computation using the explicit representation \eqref{eq:C6}, we find
\begin{align}
\braket{F(\mathcal{A})\wedge F(\mathcal{A})}_{\beta}=&\frac{1}{4}F(\mathcal{A})^{IJ}\wedge F(\mathcal{A})^{KL}\braket{M_{IJ},M_{KL}}_{\beta}-F(\mathcal{A})_r^{\alpha}\wedge F(\mathcal{A})_s^{\delta}\braket{Q^r_{\alpha},Q^s_{\delta}}_{\beta}\nonumber\\
&-\frac{1}{2L^2}\hat{F}\wedge\mathbb{P}_{\beta}\hat{F}\nonumber\\
=&\braket{F(\omega)\wedge F(\omega)}_{\beta}+\frac{2}{L^2}\braket{\Sigma\wedge F(\omega)}_{\beta}-\frac{1}{2L}\braket{F(\omega)\wedge\bar{\Psi}^r\wedge\gamma^{\bullet\bullet}\Psi_r}_{\beta}\nonumber\\
&-\frac{1}{2L^3}\braket{\Sigma\wedge\bar{\Psi}^r\wedge\gamma^{\bullet\bullet}\Psi_r}_{\beta}+\frac{1}{L^4}\braket{\Sigma\wedge\Sigma}_{\beta}+\frac{1}{L}F(\mathcal{A})_r^{\alpha}\wedge F(\mathcal{A})_s^{\delta}\delta^{rs}(C\mathcal{P}_{\beta})_{\alpha\delta}\nonumber\\
&-\frac{1}{4L^2}\hat{F}\wedge\star \hat{F}-\frac{1}{4\beta L^2}\hat{F}\wedge \hat{F}
\label{eq:6.9}
\end{align}
where
\begin{equation}
\hat{F}\wedge \hat{F}=\mathrm{d}\hat{A}\wedge\mathrm{d}\hat{A}+\mathrm{d}\hat{A}\wedge\bar{\Psi}^r\wedge\Psi^s\epsilon_{rs}+\frac{1}{4}\bar{\Psi}^r\wedge\Psi^s\epsilon_{rs}\wedge\bar{\Psi}^p\wedge\Psi^q\epsilon_{pq}
\label{eq:6.10}
\end{equation}
Let us further expand the terms in \eqref{eq:6.9} arising form the Lorentzian components of the curvature which gives
\begin{align}
&\braket{F(\omega)\wedge F(\omega)}_{\beta}+\frac{2}{L^2}\braket{\Sigma\wedge F(\omega)}_{\beta}+\frac{i}{8L}F(\omega)^{IJ}\wedge\bar{\Psi}^r\wedge\gamma_{*}\gamma^{KL}\mathcal{P}_{\beta}\Psi_r\epsilon_{IJKL}\nonumber\\
&+\frac{i}{8L^3}\Sigma^{IJ}\wedge\bar{\Psi}^r\wedge\gamma_{*}\gamma^{KL}\mathcal{P}_{\beta}\Psi_r\epsilon_{IJKL}+\frac{1}{8L^4}\Sigma^{IJ}\wedge\Sigma^{KL}\epsilon_{IJKL}\nonumber\\
&+\frac{1}{32L^2}\bar{\Psi}^r\wedge\gamma^{IJ}\Psi_r\wedge\bar{\Psi}^s\wedge\gamma_{IJ}\mathcal{P}_{\beta}\Psi_s
\label{eq:6.11}
\end{align}
In contrast to the $\mathcal{N}=1$ case, an additional $\Psi^4$-order term appears which, in general, no longer vanishes since the supermultiplet contains two independent Majorana fermions. In order to further evaluate this term, let us split the fermionic fields in their chiral components and use the Fierz-identities \eqref{eq:B1.5a}-\eqref{eq:B1.6}. In this way, it follows (summation over repeated indices)

\begin{align}
\bar{\Psi}^r\wedge\gamma^{IJ}\Psi_r\wedge\bar{\Psi}^s\wedge\gamma_{IJ}\Psi_s=&\bar{\psi}^r\wedge\gamma^{IJ}\psi^r\wedge\bar{\psi}^s\wedge\gamma_{IJ}\psi^s\nonumber\\
&+\bar{\psi}_r\wedge\gamma^{IJ}\psi_r\wedge\bar{\psi}_s\wedge\gamma_{IJ}\psi_s\nonumber\\
=&4\bar{\psi}^r\wedge\psi^s\epsilon_{rs}\wedge\bar{\psi}^p\wedge\psi^q\epsilon_{pq}\nonumber\\
&+4\bar{\psi}_r\wedge\psi_s\epsilon^{rs}\wedge\bar{\psi}_p\wedge\psi_q\epsilon^{pq}
\label{eq:6.12}
\end{align}
On the other hand, we have
\begin{align}
\bar{\Psi}^r\wedge\gamma^{IJ}\Psi_r\wedge\bar{\Psi}^s\wedge\gamma_{IJ}\gamma_{*}\Psi_s=&\bar{\psi}^r\wedge\gamma^{IJ}\psi^r\wedge\bar{\psi}^s\wedge\gamma_{IJ}\psi^s\nonumber\\
&-\bar{\psi}_r\wedge\gamma^{IJ}\psi_r\wedge\bar{\psi}_s\wedge\gamma_{IJ}\psi_s\nonumber\\
=&4\bar{\Psi}^r\wedge\Psi^s\epsilon_{rs}\wedge\bar{\Psi}^p\wedge\gamma_{*}\Psi^q\epsilon_{pq}
\label{eq:6.13}
\end{align}
Thus, using \eqref{eq:6.12} and \eqref{eq:6.13}, we can rewrite it as follows
\begin{align}
\frac{1}{32L^2}\bar{\Psi}^r\wedge\gamma^{IJ}\Psi_r\wedge\bar{\Psi}^s\wedge\gamma_{IJ}\mathcal{P}_{\beta}\Psi_s=&\frac{i}{16L^2}\bar{\Psi}^r\wedge\Psi^s\epsilon_{rs}\wedge\bar{\Psi}^p\wedge\gamma_{*}\Psi^q\epsilon_{pq}\nonumber\\
&+\frac{1}{16\beta L^2}(\bar{\psi}^r\wedge\psi^s\epsilon_{rs}\wedge\bar{\psi}^p\wedge\psi^q\epsilon_{pq}\nonumber\\
&+\bar{\psi}_r\wedge\psi_s\epsilon^{rs}\wedge\bar{\psi}_p\wedge\psi_q\epsilon^{pq})
\label{eq:6.14}
\end{align}
Hence, if we combine this with the $\beta$-dependent $\hat{F}\wedge \hat{F}$-term in the expansion \eqref{eq:6.9} given by the expression \eqref{eq:6.10}, this yields
\begin{align}
-&\frac{1}{4\beta L^2}\hat{F}\wedge \hat{F}+\frac{1}{32L^2}\bar{\Psi}^r\wedge\gamma^{IJ}\Psi_r\wedge\bar{\Psi}^s\wedge\gamma_{IJ}\mathcal{P}_{\beta}\Psi_s\nonumber\\
=&-\frac{1}{4\beta L^2}\mathrm{d}(\hat{A}\wedge\mathrm{d}\hat{A})-\frac{1}{4\beta L^2}\mathrm{d}\hat{A}\wedge\bar{\Psi}^r\wedge\Psi^s\epsilon_{rs}+\frac{i}{16L^2}\bar{\Psi}^r\wedge\Psi^s\epsilon_{rs}\wedge\bar{\Psi}^p\wedge\gamma_{*}\Psi^q\epsilon_{pq}\nonumber\\
&-\frac{1}{8\beta L^2}\bar{\psi}^r\wedge\psi^s\epsilon_{rs}\wedge\bar{\psi}_p\wedge\psi_q\epsilon^{pq}
\label{eq:6.15}
\end{align}
The remaining terms in \eqref{eq:6.9} can be treated as in the $\mathcal{N}=1$ case. For sake of completeness, let us repeat them here. Therefore, again, notice that
\begin{align}
\frac{1}{L}F(\mathcal{A})_r^{\alpha}\wedge F(\mathcal{A})_s^{\delta}\delta^{rs}(C\mathcal{P}_{\beta})_{\alpha\delta}=&\braket{\nabla\Psi\wedge\nabla\Psi}_{\beta}-\frac{1}{L^2}\bar{\Psi}^r\wedge\boldsymbol{\gamma}\wedge\mathcal{P}_{\beta}\nabla\Psi_r\nonumber\\
&-\frac{1}{4L^3}\bar{\Psi}^r\wedge\gamma_{IJ}\mathcal{P}_{-\beta}\Psi_r\wedge\Sigma^{IJ}
\label{eq:6.16}
\end{align}
Hence, using $\tensor{\epsilon}{_{IJ}^{KL}}\gamma_{KL}=2i\gamma_{IJ}\gamma_{*}$, the last term in \eqref{eq:6.16} can be combined with the first term in the second line of \eqref{eq:6.11} to give
\begin{align}
&\frac{i}{8L^3}\Sigma^{IJ}\wedge\bar{\Psi}^r\wedge\gamma_{*}\gamma^{KL}\mathcal{P}_{\beta}\Psi_r\epsilon_{IJKL}-\frac{1}{4L^3}\bar{\Psi}^r\wedge\gamma_{IJ}\mathcal{P}_{\beta}\Psi_r\wedge\Sigma^{IJ}\nonumber\\
=&\frac{i}{8L^3}\bar{\Psi}^r\wedge\gamma_{*}\gamma^{KL}\big[\mathcal{P}_{\beta}+\mathcal{P}_{-\beta}\big]\Psi_r\wedge\Sigma^{IJ}\epsilon_{IJKL}=-\frac{1}{8L^3}\bar{\Psi}^r\wedge\gamma^{KL}\Psi_r\wedge\Sigma^{IJ}\epsilon_{IJKL}
\label{eq:6.17}
\end{align}
Thus, to summarize, it follows that the action \eqref{eq:6.4} can be written in the equivalent form
\begin{align}
S_{H-MM}(\mathcal{A})=\frac{1}{2\kappa}\int_{M}&\frac{1}{2}\Sigma^{IJ}\wedge F(\omega)^{KL}\epsilon_{IJKL}-i\bar{\Psi}^r\wedge\boldsymbol{\gamma}\gamma_{*}\wedge\nabla\Psi_r\nonumber\\
-&\frac{1}{4L}\bar{\Psi}^r\wedge\gamma^{KL}\Psi_r\wedge\Sigma^{IJ}\epsilon_{IJKL}+\frac{1}{4L^2}\Sigma^{IJ}\wedge\Sigma^{KL}\epsilon_{IJKL}\nonumber\\
+&\frac{i}{8}\bar{\Psi}^r\wedge\Psi^s\epsilon_{rs}\wedge\bar{\Psi}^p\wedge\gamma_{*}\Psi^q\epsilon_{pq}-\frac{1}{2}\hat{F}\wedge\star \hat{F}\nonumber\\
+&\frac{1}{4}F(\omega)^{IJ}\wedge F(\omega)^{KL}\epsilon_{IJKL}+iL\nabla\bar{\Psi}^r\wedge\gamma_{*}\nabla\Psi_r\nonumber\\
-&\frac{iL}{4}F(\omega)^{IJ}\wedge\bar{\Psi}^r\wedge\gamma_{*}\gamma_{IJ}\Psi_r+\frac{1}{\beta}\mathscr{L}_{\beta}
\label{eq:6.18}
\end{align}
where we have collected all terms depending on the Barbero-Immirzi parameter in the Lagrangian $\mathscr{L}_{\beta}$ given by
\begin{align}
\mathscr{L}_{\beta}=&\frac{1}{2}F(\omega)^{IJ}\wedge F(\omega)_{IJ}+\Sigma^{IJ}\wedge F(\omega)_{IJ}-\frac{1}{4}\bar{\psi}^r\wedge\psi^s\epsilon_{rs}\wedge\bar{\psi}_p\wedge\psi_q\epsilon^{pq}-\bar{\Psi}^r\wedge\boldsymbol{\gamma}\wedge\nabla\Psi_r\nonumber\\
&+L\nabla\bar{\Psi}^r\wedge\nabla\Psi_r-\frac{L}{4}F(\omega)^{IJ}\wedge\bar{\Psi}^r\wedge\gamma_{IJ}\Psi_r-\frac{1}{2}\mathrm{d}\hat{A}\wedge\bar{\Psi}^r\wedge\Psi^s\epsilon_{rs}-\frac{1}{2}\mathrm{d}(\hat{A}\wedge\mathrm{d}\hat{A})
\label{eq:6.19}
\end{align}
We have to show that this Lagrangian is indeed topological at second order, i.e. it takes the form of a boundary term provided the spin connection satisfies its field equations. For the second line in \eqref{eq:6.19}, this is an immediate consequence of the identity
\begin{align}
L\nabla\bar{\Psi}^r\wedge\nabla\Psi_r=&\mathrm{d}(L\bar{\Psi}^r\wedge\nabla\Psi_r)+\bar{\Psi}^r\wedge\nabla\nabla\Psi_r\nonumber\\
=&\mathrm{d}(L\bar{\Psi}^r\wedge\nabla\Psi_r)+\frac{L}{4}F(\omega)^{IJ}\wedge\bar{\Psi}^r\wedge\gamma_{IJ}\Psi_r+\frac{1}{2}\mathrm{d}\hat{A}\wedge\bar{\Psi}^r\wedge\Psi^s\epsilon_{rs}
\label{eq:6.20}
\end{align}
For the first line, note that the EOM of $\omega$ are equivalent to the supertorsion constraint $F(\mathcal{A})^I=0$, that is
\begin{align}
D^{(\omega)}e^I\equiv\Theta^{(\omega)I}=\frac{1}{4}\bar{\Psi}^r\wedge\gamma^I\Psi_r
\label{eq:6.21}
\end{align}
Thus, using \eqref{eq:6.21}, we can rewrite the last term in the first line of \eqref{eq:6.19} as follows
\begin{align}
\bar{\Psi}^r\wedge\boldsymbol{\gamma}\wedge\nabla\Psi_r=&\frac{1}{2}\mathrm{d}(\bar{\Psi}^r\wedge\boldsymbol{\gamma}\wedge\Psi_r)+\frac{1}{2}\bar{\Psi}^r\wedge D^{(\omega)}e^I\gamma_I\wedge\Psi_r\nonumber\\
=&\frac{1}{2}\mathrm{d}(\bar{\Psi}^r\wedge\boldsymbol{\gamma}\wedge\Psi_r)+\frac{1}{8}\bar{\Psi}^r\wedge\gamma_I\Psi_r\wedge\bar{\Psi}^s\wedge\gamma^I\Psi_s\nonumber\\
=&\frac{1}{2}\mathrm{d}(\bar{\Psi}^r\wedge\boldsymbol{\gamma}\wedge\Psi_r)+\frac{1}{2}\bar{\psi}_r\wedge\gamma_I\wedge\psi^r\wedge\bar{\psi}^s\wedge\gamma^I\wedge\psi_s\nonumber\\
=&\frac{1}{2}\mathrm{d}(\bar{\Psi}^r\wedge\boldsymbol{\gamma}\wedge\Psi_r)-\frac{1}{2}\bar{\psi}^r\wedge\psi^s\epsilon_{rs}\wedge\bar{\psi}_p\wedge\psi_q\epsilon^{pq}
\label{eq:6.22}
\end{align}
On the other hand, according to \eqref{eq:B1.5a}-\eqref{eq:B1.6}, we have the important identity
\begin{align}
\Theta^{(\omega)I}\wedge\Theta^{(\omega)}_I&=\frac{1}{16}\bar{\Psi}^r\wedge\gamma_I\Psi_r\wedge\bar{\Psi}^s\wedge\gamma^I\Psi_s=-\frac{1}{4}\bar{\psi}^r\wedge\psi^s\epsilon_{rs}\wedge\bar{\psi}_p\wedge\psi_q\epsilon^{pq}
\label{eq:6.23}
\end{align}
Hence, the last three terms in the first line of \eqref{eq:6.19} can be re-expressed in the following way
\begin{align}
&\Sigma^{IJ}\wedge F(\omega)_{IJ}-\frac{1}{4}\bar{\psi}^r\wedge\psi^s\epsilon_{rs}\wedge\bar{\psi}_p\wedge\psi_q\epsilon^{pq}-\bar{\Psi}^r\wedge\boldsymbol{\gamma}\wedge\nabla\Psi_r\nonumber\\
=&\Sigma^{IJ}\wedge F(\omega)_{IJ}-\Theta^{(\omega)I}\wedge\Theta^{(\omega)}_I+2\mathrm{d}(e^I\wedge\Theta^{(\omega)}_I)
\label{eq:6.24}
\end{align}
In fact, this can be simplified even further. Therefore, we notice that the first two terms in equation \eqref{eq:6.24} yield the so-called topological \emph{Nieh-Yan term} $\mathrm{d}(e^I\wedge\Theta^{(\omega)}_I)$ \cite{Nieh:1981ww}. This is easy to see using the properties of the covariant derivative which immediately gives\footnote{This can also be checked by direct computation:
\begin{align*}
\mathrm{d}(e^I\wedge\Theta^{(\omega)}_I)&=\mathrm{d}e^I\wedge\Theta^{(\omega)}_I-e^I\wedge\mathrm{d}\Theta^{(\omega)}_I=\mathrm{d}e^I\wedge\Theta^{(\omega)}_I+\tensor{\omega}{^J_I}\wedge e^I\wedge\mathrm{d}e_J-e^I\wedge e^J\wedge\mathrm{d}\omega_{IJ}\nonumber\\
&=\mathrm{d}e^I\wedge\Theta^{(\omega)}_I+\tensor{\omega}{^J_I}\wedge e^I\wedge\mathrm{d}e_J+\tensor{\omega}{^K_J}\wedge e^J\wedge\tensor{\omega}{_K^I}\wedge e_I-e^I\wedge e^J\wedge F(\omega)_{IJ}\nonumber\\
&=\Theta^{(\omega)I}\wedge\Theta^{(\omega)}_I-\Sigma^{IJ}\wedge F(\omega)_{IJ}
\end{align*}
 }
\begin{align}
\mathrm{d}(e^I\wedge\Theta^{(\omega)}_I)&=D^{(\omega)}e^I\wedge\Theta^{(\omega)}_I-e^I\wedge D^{(\omega)}D^{(\omega)}e_I\nonumber\\
&=\Theta^{(\omega)I}\wedge\Theta^{(\omega)}_I-\Sigma^{IJ}\wedge F(\omega)_{IJ}
\label{eq:6.25}
\end{align}
Thus, to summarize, we see, provided the spin connection satisfies its field equations, the Lagrangian \eqref{eq:6.19} takes the final form  
\begin{align}
\mathscr{L}_{\beta}=&\frac{1}{2}F(\omega)^{IJ}\wedge F(\omega)_{IJ}+\mathrm{d}\!\left(e^I\wedge\Theta^{(\omega)}_I+L\bar{\Psi}^r\wedge\nabla\Psi_r-\frac{1}{2}\hat{A}\wedge\mathrm{d}\hat{A}\right)
\label{eq:6.26}
\end{align}
and therefore is indeed topological. Moreover, if we subtract this term from the full action \eqref{eq:6.4}, it follows that this action finally reduces to
\begin{align}
S(\mathcal{A})=\frac{1}{2\kappa}\int_{M}&\frac{1}{2}\Sigma^{IJ}\wedge F(\omega)^{KL}\epsilon_{IJKL}-i\bar{\Psi}^r\wedge\boldsymbol{\gamma}\gamma_{*}\wedge\nabla\Psi_r\nonumber\\
-&\frac{1}{4L}\bar{\Psi}^r\wedge\gamma^{KL}\Psi_r\wedge\Sigma^{IJ}\epsilon_{IJKL}+\frac{1}{L^2}\Sigma^{IJ}\wedge\Sigma^{KL}\epsilon_{IJKL}\nonumber\\
+&\frac{i}{2}\left(\mathrm{d}\hat{A}+\frac{1}{4}\bar{\Psi}^r\wedge\Psi^s\epsilon_{rs}\right)\wedge\bar{\Psi}^p\wedge\gamma_{*}\Psi^q\epsilon_{pq}-\frac{1}{2}\hat{F}\wedge\star \hat{F}+\mathscr{L}_{\text{bdy}}
\label{eq:6.27}
\end{align}
with $\mathscr{L}_{\text{bdy}}$ given by \eqref{eq:5.8} (times $2L^{-2}$). Hence, at second order, the Holst action leads back to the original action of $\mathcal{N}=2$, $D=4$ AdS supergravity as stated in \cite{Andrianopoli:2014aqa,Andrianopoli:2020zbl} as required. To summarize, the Holst-MacDowell-Mansouri action can be written in the form
\begin{align}
S_{H-MM}(\mathcal{A})=\frac{1}{2\kappa}\int_{M}&\Sigma^{IJ}\wedge (P_{\beta}\circ F(\omega))^{KL}\epsilon_{IJKL}-\bar{\Psi}^r\wedge\boldsymbol{\gamma}\frac{\mathds{1}+i\beta\gamma_{*}}{\beta}\wedge\nabla\Psi_r\nonumber\\
-&\frac{1}{4L}\bar{\Psi}^r\wedge\gamma^{KL}\Psi_r\wedge\Sigma^{IJ}\epsilon_{IJKL}+\frac{1}{L^2}\Sigma^{IJ}\wedge\Sigma^{KL}\epsilon_{IJKL}\nonumber\\
+&\frac{i}{2}\left(\mathrm{d}\hat{A}+\frac{1}{4}\bar{\Psi}^r\wedge\Psi^s\epsilon_{rs}\right)\wedge\bar{\Psi}^p\wedge\gamma_{*}\Psi^q\epsilon_{pq}-\frac{1}{4\beta}\bar{\psi}^r\wedge\psi^s\epsilon_{rs}\wedge\bar{\psi}_p\wedge\psi_q\epsilon^{pq}\nonumber\\
-&\frac{1}{2}\hat{F}\wedge\star \hat{F}+S_{\text{bdy}}
\label{eq:6.27}
\end{align}
with $S_{\text{bdy}}$ a boundary action given by 
\begin{align}
    S_{\text{bdy}}(\mathcal{A})=\frac{L^2}{\kappa}\int_{\partial M}{\braket{\omega\wedge\mathrm{d}\omega+\frac{1}{3}\omega\wedge[\omega\wedge\omega]}_{\beta}-\frac{1}{4\beta L^2}\hat{A}\wedge\mathrm{d}\hat{A}+\braket{\Psi\wedge\nabla\Psi}_{\beta}}
    \label{eq:bdyN2}
\end{align}
In particular, according to the general discussion in section \ref{Section:Pure N=2}, similar to the $\mathcal{N}=1$ case, this boundary action is determined uniquely if one requires supersymmetry invariance of the full action at the boundary. 

Again, as in the non-extended case, since the inner product in \eqref{eq:bdyN2} is not invariant under the full supergroup $\mathrm{OSp}(2|4)$ but only under the action of its bosonic subgroup, the boundary action \eqref{eq:bdyN2}, in general, will not correspond to a super Chern-Simons action. As we will see in the following section, this changes however in case of the chiral theory where the boundary theory will generically be described in terms of a super Chern-Simons theory with gauge supergroup $\mathrm{OSp}(2|2)_{\mathbb{C}}$.

Nevertheless, one should emphasize that, at least in context of the standard underformed theory, one can construct models where this turns out to be true even in case of classical (real) variables. For instance, in \cite{Andrianopoli:2018ymh} particular falloff conditions for the physical fields in the $\mathcal{N}=2$ case where considered leading to a super Chern-Simons theory on the boundary corresponding to a $\mathrm{OSp}(2|2)\times\mathrm{SO}(1,2)$ gauge group. This model has also been studied in \cite{Alvarez:2011gd,Guevara:2016rbl,Alvarez:2013tga} which turned out to have interesting applications in condensed matter physics in the description of graphene near the Dirac points.

\begin{remark}
It is interesting to note that, via definition \eqref{eq:6.1} and \eqref{eq:6.3}, the Barbero-Immirzi parameter leads to an additional topological term in the $\mathrm{U}(1)$ sector of the theory which is also known as the $\theta$-term in Yang-Mills theory. Hence, in this sense, the Barbero-Immirzi parameter literally has the interpretation in terms of a $\theta$-ambiguity. This supports the hypothesis of \cite{Obregon:2012zz}. It could also have consequences for the quantum theory of the U(1) sector, but this is beyond the scope of the present work.  
\end{remark}
%First, notice that 
%\begin{align}
%&\frac{2}{L^2}\braket{\Sigma\wedge F(\omega)}_{\beta}=\frac{1}{2L^2}\Sigma^{IJ}\wedge F(\omega)^{KL}\epsilon_{IJKL}+\frac{1}{\beta L^2}\Sigma^{IJ}\wedge F(\omega)_{IJ}
%\label{eq:}
%\end{align}
\subsection{The chiral theory}\label{Section:Chiral N=2}
Let us finally consider the chiral limit setting $\beta=-i$ for the Barbero-Immirzi parameter. Then, the operator \eqref{eq:6.1} takes the form $\mathbf{P}_{-i}=i\mathbf{P}^+$ with 
\begin{equation}
\mathbf{P}^+:\,\Omega^2(M,\mathfrak{osp}(2|4)_{\mathbb{C}})\rightarrow\Omega^2(M,\mathfrak{osp}(2|2)_{\mathbb{C}})
\label{eq:7.1}
\end{equation}
the projection onto differential forms on the bosonic spacetime manifold with values in the orthosymplectic subalgebra $\mathfrak{osp}(2|2)_{\mathbb{C}}$. In order to see the underlying $\mathrm{OSp}(2|2)_{\mathbb{C}}$-gauge symmetry of the theory, let us introduce the \emph{super Ashtekar connection} $\mathcal{A}^+$ by projection the super Cartan connection \eqref{eq:5.1} onto the chiral subalgebra yielding
%In contrast to the $\mathcal{N}=1$ case, when inserting this projection into ??, the action does not appear manifestly $\mathrm{OSp}(2|2)_{\mathbb{C}}$-invariant. This is due to the fact that the projection ?? does not commute with the Adjoint representation of the chiral super subgroup which is due to the appearence of the Hodge-dualization. This, on the other hand, results from the fact that we are working at second order in the $\mathrm{U}(1)$ gauge field as the Hodge-star operation cannot directly be defined on the supermanifold. However, as it will become clear in what follows, it is still possible
\begin{equation}
    \mathcal{A}^+:=\mathbf{P}^{\mathfrak{osp}(2|2)}\mathcal{A}=A^{+i}T^+_i+AT+\psi^A_rQ^r_A
    \label{eq:7.2}
\end{equation}
which, as can be checked directly, again defines a generalized super Cartan connection and therefore yields a proper connection on the associated $\mathrm{OSp}(2|2)_{\mathbb{C}}$-bundle. Applying the projection on the super curvature, we obtain for the Lorentzian sub components\footnote{We stick to our notation and write $\psi^A_r$ and $\bar{\psi}_{A'}^r$ for the chiral and anti chiral components of the Majorana fermion fields, respectively. The position of the $R$-symmetry index for the chiral components stays fixed. Moreover, we will sum over repeated indices.}  
\begin{equation}
    (\mathbf{P}^{\mathfrak{osp}(2|2)}F(\mathcal{A}))^i=F(A^+)^i+\frac{i}{L}\psi^A_r\wedge\psi^B_r\tau^i_{AB}+\frac{1}{2L^2}\Sigma^i=F(\mathcal{A}^+)^i+\frac{1}{2L^2}\Sigma^i
    \label{eq:7.3}
\end{equation}
with $F(\mathcal{A}^+)$ the curvature of $\mathcal{A}^+$ and $\Sigma^i$ defined as in the $\mathcal{N}=1$ case. Moreover, for the chiral odd components, we find 
\begin{equation}
(\mathbf{P}^{\mathfrak{osp}(2|2)}F(\mathcal{A}))^A_r=D^{(A^+)}\psi^A_r+\frac{1}{2L}A\epsilon_{rs}\wedge\psi^A_s+\frac{1}{2L}\chi^A_r=F(\mathcal{A}^+)^A_r+\frac{1}{2L}\chi^A_r
\label{eq:7.4}
\end{equation}
where $\chi^A_r=-e^{AA'}\wedge\bar{\psi}_{A'}^s\delta_{rs}$. Finally, for the $\mathrm{U}(1)$-component, we get
\begin{equation}
    \mathbf{P}^{\mathfrak{osp}(2|2)}\hat{F}=\hat{F}=\hat{F}^++\frac{i}{2}\bar{\psi}^r_{A'}\wedge\bar{\psi}^{A's}\epsilon_{rs}
    \label{eq:7.5}
\end{equation}
with $\hat{F}^+:=\mathrm{d}A+\frac{i}{2}\psi_A^r\wedge\psi^{As}\epsilon_{rs}$. To summarize, we can decompose the super Cartan curvature in the following way
\begin{equation}
    \mathbf{P}^{\mathfrak{osp}(2|2)}F(\mathcal{A})=:F(\mathcal{A}^+)+\frac{1}{2L^2}\tilde{\mathcal{E}}
    \label{eq:7.6}
\end{equation}
where $\tilde{\mathcal{E}}$ is a graded field which (in constrast to $\mathcal{E}$ to be defined below), as we would like to emphasize, does not have a simple transformation behavior as in the $\mathcal{N}=1$ case under left-handed supersymmetry transformations. This is due to the fact that, for $\mathcal{N}=2$ and in contrast to the pure $\mathcal{N}=1$ case, supersymmetry transformations which generically have to regarded as superdiffeomorphisms no longer have an equivalent characterization in terms of super gauge transformations leading to nontrivial curvature contributions in the SUSY variations of the super Cartan connection according to the general formula \eqref{eq:5.7}.\\
Let us insert \eqref{eq:7.6} into \eqref{eq:6.4} for $\beta=-i$ which gives the \emph{chiral Holst-MacDowell-Mansouri action}
\begin{equation}
S(\mathcal{A})=\frac{iL^2}{\kappa}\int_{M}{\braket{[F(\mathcal{A}^+)+\frac{1}{2L^2}\tilde{\mathcal{E}}]\wedge\mathbf{P}^+[F(\mathcal{A}^+)+\frac{1}{2L^2}\tilde{\mathcal{E}}]}}
\label{eq:7.6.1}
\end{equation}
For reasons that will become clear in a moment, let us next subtract the topological term $iL^2\braket{F(\mathcal{A}^+)\wedge F(\mathcal{A}^+)}/\kappa$ of the full action \eqref{eq:7.6.1}. If we then define the projection $\mathbb{P}^-:=\underline{\mathbf{0}}\oplus\frac{1}{2}(1+i\star)\oplus\underline{\mathbf{0}}$ projecting onto the anti self-dual part of the $\mathrm{U}(1)$-sub component of $F(\mathcal{A}^+)$, it follows that the bulk contribution in \eqref{eq:7.6.1} takes the form 
\begin{align}
S_{\text{bulk}}(\mathcal{A})&=\frac{i}{\kappa}\int_{M}{-L^2\braket{F(\mathcal{A}^+)\wedge\mathbb{P}^-F(\mathcal{A}^+)}+\braket{F(\mathcal{A^+})\wedge\mathbf{P}^+\tilde{\mathcal{E}}}+\frac{1}{4L^2}\braket{\tilde{\mathcal{E}}\wedge\mathbf{P}^+\tilde{\mathcal{E}}}}\nonumber\\
&=\frac{i}{\kappa}\int_{M}{\braket{F(\mathcal{A^+})\wedge[\mathbf{P}^+\tilde{\mathcal{E}}-L^2\mathbb{P}^-F(\mathcal{A}^+)]}+\frac{1}{4L^2}\braket{\tilde{\mathcal{E}}\wedge\mathbf{P}^+\tilde{\mathcal{E}}}}
\label{eq:7.7}
\end{align}
As it turns out, \eqref{eq:7.7} can be rewritten in a very intriguing form. In fact, let us define $\mathcal{E}:=\mathbf{P}^+\tilde{\mathcal{E}}-2L^2\mathbb{P}^-F(\mathcal{A}^+)$ for the \emph{super electric field}. It then follows that the bulk action \eqref{eq:7.7} is equivalent to
\begin{align}
S_{\text{bulk}}(\mathcal{A})=\frac{i}{\kappa}\int_{M}{\braket{F(\mathcal{A^+})\wedge\mathcal{E}}+\frac{1}{4L^2}\braket{\mathcal{E}\wedge\mathcal{E}}}
\label{eq:7.8}
\end{align}
This follows immediately from the fact that both $\mathbf{P}^+$ and $\mathbb{P}^-$ define projections projecting onto mutually orthogonal subspaces such that $\mathbf{P}^+\circ\mathbb{P}^-=0=\mathbb{P}^-\circ\mathbf{P}^+$ which yields $\braket{\mathcal{E}\wedge\mathcal{E}}=\braket{\tilde{\mathcal{E}}\wedge\mathbf{P}^+\tilde{\mathcal{E}}}+4L^4\braket{F(\mathcal{A}^+)\wedge\mathbb{P}^-F(\mathcal{A}^+)}$. Hence, the bulk action takes the form of a Palatini-type action with nontrivial cosmological constant written in chiral variables and   $\mathrm{OSp}(2|2)_{\mathbb{C}}$ structure group. It is interesting to note that the subtraction of the CS-topological term from the full action was crucial for this result leading to the projection $\mathbb{P}^-$ which is orthogonal to the chiral projection $\mathbf{P}^+$.\\
In order to see that the super electric field $\mathcal{E}$ indeed defines the canonical conjugate of the super Ashtekar connection, let us go back to \eqref{eq:7.7} and vary the action with respect to $\mathcal{A}^+$. In this way, it follows
\begin{align}
\delta S_{\text{bulk}}(\mathcal{A})&=\frac{i}{\kappa}\int_{M}{\braket{D^{(\mathcal{A}^+)}\delta\mathcal{A^+}\wedge\mathcal{E}}}=:\mathrm{d}\Theta+\frac{i}{\kappa}\int_{M}\braket{\delta\mathcal{A}^+\wedge D^{(\mathcal{A}^+)}\mathcal{E}}
\label{eq:7.9}
\end{align}
with pre-symplectic potential $\Theta(\delta)$ inducing the bulk pre-symplectic structure
\begin{equation}
    \Omega_{\text{bulk}}(\delta_1,\delta_2)=\frac{2i}{\kappa}\int_{\Sigma}\braket{\delta_{[1}\mathcal{A}^+\wedge\delta_{2]}\mathcal{E}}
    \label{eq:7.10}
\end{equation}
Thus, indeed, $(\mathcal{A}^+,\mathcal{E})$ define the fundamental variables of the canonical phase space. Moreover, from \eqref{eq:7.9}, we deduce that the Gauss constraint $\mathcal{G}(\Lambda)$, for any smooth $\mathfrak{osp}(2|2)_{\mathbb{C}}$-valued smearing function $\Lambda$, takes the form
\begin{equation}
    \mathcal{G}(\Lambda)=\int_{\Sigma}\braket{D^{(\mathcal{A}^+)}\mathcal{E},\Lambda}
    \label{eq:7.11}
\end{equation}
and, as a consequence, satisfies the constraint algebra $\{\mathcal{G}(\Lambda),\mathcal{G}(\Lambda')\}=\mathcal{G}([\Lambda,\Lambda'])$. That is, the Gauss constraint generates local $\mathrm{OSp}(2|2)_{\mathbb{C}}$-gauge transformations.\\
The boundary action of the theory takes the form 
\begin{equation}
S_{\text{bdy}}(\mathcal{A}^+)=\frac{iL^2}{\kappa}\int_{M}{\braket{F(\mathcal{A}^+)\wedge F(\mathcal{A}^+)}}=\frac{iL^2}{\kappa}\int_{\partial M}{\braket{\mathcal{A}^+\wedge\mathrm{d}\mathcal{A}^++\frac{1}{3}\mathcal{A}^+\wedge[\mathcal{A}^+\wedge\mathcal{A}^+]}}
\label{eq:7.12}
\end{equation}
and thus, in particular, corresponds to the action of a $\mathrm{OSp}(2|2)_{\mathbb{C}}$ super Chern-Simons theory with (complex) Chern-Simons level $k=i4\pi L^2/\kappa=-i12\pi/\kappa\Lambda$ with $\Lambda$ the cosmological constant. The pre-symplectic structure of the full theory is given by
\begin{equation}
    \Omega(\delta_1,\delta_2)=\frac{2i}{\kappa}\int_{\Sigma}\braket{\delta_{[1}\mathcal{A}^+\wedge\delta_{2]}\mathcal{E}}-\frac{k}{2\pi}\int_{\partial\Sigma}\braket{\delta_{[1}\mathcal{A}^+\wedge\delta_{2]}\mathcal{A}^+}
    \label{eq:7.13}
\end{equation}
As in the $\mathcal{N}=1$ case, due to the splitting of the full action into a bulk and boundary term, one needs to derive a matching condition relating bulk and boundary degrees of freedom at the boundary. This is equivalent to requiring consistency with the equation of motion of the full theory , i.e. $\delta S=\delta S_{\text{bulk}}+\delta S_{\text{bdy}}=0$. From this we can immediately read off the boundary condition
\begin{equation}
    \underset{\raisebox{1pt}{$\Leftarrow$}}{\mathcal{E}}=\frac{i\kappa k}{2\pi}\underset{\raisebox{1pt}{$\Longleftarrow$}}{F(\mathcal{A}^{+})}
    \label{eq:7.14}
\end{equation}
where, again, the arrow denotes the pullback to the boundary. This condition relates the super electric field $\mathcal{E}$ to the curvature of the super connection $\mathcal{A}^+$ corresponding to the $\mathrm{OSp}(2|2)_{\mathbb{C}}$ super Chern-Simons theory living on the boundary. 
\begin{remark}
Note that boundary condition \eqref{eq:7.14} can equivalently be rewritten in the following form
\begin{equation}
    \underset{\raisebox{1pt}{$\Longleftarrow$}}{F(\mathcal{A}^{+})}+\frac{1}{2L^2}\underset{\raisebox{1pt}{$\Leftarrow$}}{\mathcal{E}}=0\quad\Leftrightarrow\quad\underset{\raisebox{1pt}{$\Longleftarrow$}}{\mathbf{P}^+F(\mathcal{A})}=0
    \label{eq:7.15}
\end{equation}
where we used that the $\mathrm{U}(1)$-component $\hat{\mathcal{E}}$ of the super electric field can be written as $\hat{\mathcal{E}}=\hat{\tilde{\mathcal{E}}}-L^2(1+i\star)\hat{F}$ such that
\begin{equation}
    \hat{F}^++\frac{1}{2L^2}\hat{\mathcal{E}}=\hat{F}-\frac{1}{2}(1+i\star)\hat{F}=\frac{1}{2}(1-i\star)\hat{F}
    \label{eq:7.16}
\end{equation}
Thus, if we take the complex conjugate of \eqref{eq:7.15} yielding $\mathbf{P}^-F(\mathcal{A})=0$, we find that condition \eqref{eq:7.14} is equivalent to
\begin{equation}
    \underset{\raisebox{1pt}{$\Longleftarrow$}}{\mathbf{P}^+F(\mathcal{A})}+\underset{\raisebox{1pt}{$\Longleftarrow$}}{\mathbf{P}^-F(\mathcal{A})}=0%\quad\Leftrightarrow\quad \underset{\raisebox{1pt}{$\Longleftarrow$}}{F(\mathcal{A})}^{IJ}=0,\quad\underset{\raisebox{1pt}{$\Longleftarrow$}}{F(\mathcal{A})}^{\alpha}_r=0\text{ and }\underset{\raisebox{1pt}{$\Leftarrow$}}{\hat{F}}=0
    \label{eq:7.17}
\end{equation}
that is, the pullback of the curvature components $F(\mathcal{A})^{IJ}$, $F(\mathcal{A})^{\alpha}_r$ and $\hat{F}$ corresponding to the $\mathrm{OSp}(2|4)$ super Cartan connection $\mathcal{A}$ to the boundary are constrained to vanish at the boundary in accordance with the boundary condition as derived in \cite{Andrianopoli:2014aqa} in context of the non-chiral theory.  
\end{remark}

\section{Discussion and outlook}
In this present article, we have addressed several questions concerning the classical description of (extended) supergravity theories in $D=4$ in terms of a new type of action we called Holst-MacDowell-Mansouri action, in the presence of boundaries. For a special choice of the Barbero-Immirzi parameter, one obtains a description in terms of chiral Ashtekar type variables in which case the theory has many interesting properties and which seem to be of particular relevance for applications in LQG. In particular, we considered the question of how to properly include boundary terms in the theory. This is crucial as the standard treatment of inner boundaries in (super)gravity theories expressed in terms of (real) Ashtekar-Barbero variables is based on the isolated horizon (IH) formalism which, so far, does not take into account local supersymmetry invariance at the boundary.\footnote{In fact, the standard boundary conditions arising in this formalism even seem to break local supersymmetry.}

Hence, in this article, we have followed a different route using new developments in the geometric approach to supergravity. More precisely, following \cite{Andrianopoli:2014aqa,Andrianopoli:2020zbl}, we have discussed the most general ansatz of possible boundary terms to be added to the bulk action of $\mathcal{N}=1$ and $\mathcal{N}=2$ pure AdS supergravity in $D=4$. \cite{Andrianopoli:2014aqa,Andrianopoli:2020zbl} show that the boundary terms are fixed uniquely if one requires invariance of the full action under supersymmetry transformations at the boundary. Moreover, it follows that the resulting action in both cases acquires a very intriguing form extending the well-known MacDowell-Mansouri action \cite{MacDowell:1977jt} even to supergravity theories with extended supersymmetry \cite{Andrianopoli:2014aqa,Andrianopoli:2020zbl}.

Based on these results, we have derived a Holst variant of the MacDowell-Masouri action including topological terms for arbitrary Barbero-Immirzi paramters $\beta$ for the cases $\mathcal{N}=1,2$. To this end, inspired by ideas of \cite{Randono:2006ru,Wise:2009fu,Freidel:2005ak} in context of ordinary first order Einstein gravity, we introduced a $\beta$-deformed inner product defined via a quasi-projection operator $\mathbf{P}_{\beta}$ acting on super Lie algebra-valued forms. We have then shown that the resulting action is indeed independent of the Barbero-Immirzi parameter at second order, i.e. provided the spin connection satisfies its field equations, in the sense that all $\beta$-dependent terms become purely topological.

Moreover, for this result to be true, in case $\mathcal{N}=2$, we have seen that this required the inclusion of an additional $\beta$-dependent topological term to the Maxwell-kinetic term in the Lagrangian corresponding to the graviphoton field which is commonly known as a $\theta$-term in Yang-Mills theory. Hence, this supports the hypothesis as discussed e.g. in \cite{Obregon:2012zz}, that the Barbero-Immirzi parameter has to be regarded as kind of a $\theta$-ambiguity.

We have then studied the boundary terms arising from the Holst action. However, these boundary terms in general turn out to not correspond to a (super) Chern-Simons theory. This is, of course, in contrast to the results in context of ordinary gravity studying non-supersymmetric isolated horizon (IH) boundary conditions with Ashtekar-Barbero variables. There, one finds that, generically, the boundary theory is described via Chern-Simons theories. Nevertheless, one should emphasize that one can construct models where this turns out to be true even in the supersymmetric case. For instance, in \cite{Andrianopoli:2018ymh}, for classical variables, particular falloff conditions for the physical fields in the $\mathcal{N}=2$ case where considered leading to a super Chern-Simons theory on the boundary corresponding to a $\mathrm{OSp}(2|2)\times\mathrm{SO}(1,2)$ gauge group. This model has also been studied in \cite{Alvarez:2011gd,Guevara:2016rbl,Alvarez:2013tga} which turned out to have interesting applications in condensed matter physics in the description of graphene near the Dirac points. It is highly suggestive that similar models can also be constructed using real Ashtekar variables. This remains as a task for future investigation.

We have then turned towards the chiral limit of the theory corresponding to an imaginary $\beta=\pm i$ and have seen that the resulting theory has many interesting properties. On the one hand, it follows that the chiral action in both cases, i.e. $\mathcal{N}=1$ and $\mathcal{N}=2$, can be written in a way such that it is manifestly invariant under an enlarged gauge symmetry corresponding to the (complex) orthosymplectic group $\mathrm{OSp}(\mathcal{N}|2)_{\mathbb{C}}$. This generalizes and extends previous results obtained e.g. in \cite{Fulop:1993wi,Ezawa:1995nj,Gambini:1995db,Tsuda:2000er,Ling:2000ss}.
In particular, it follows that the boundary action takes the form a super Chern-Simons action with $\mathrm{OSp}(\mathcal{N}|2)_{\mathbb{C}}$ as a gauge group. This confirms the prescient works \cite{Smolin:1995vq,Ling:2000ss} that saw a close connection between (super) gravity in the bulk and Chern-Simons theory on the boundary.

For $\mathcal{N}=1$, we have also shown that, at the boundary, the full action is indeed invariant under both left- and right-handed supersymmetry transformations. In fact, we could even show explicitly that this requirement fixes the CS-term as the unique boundary term. In this context, we derived boundary conditions that couple bulk and boundary degrees of freedom. These turned out to be in strong similarity to the standard boundary conditions as typically considered in LQG as they imply coupling between the super electric field and the curvature of the super Ashtekar connection. Moreover, we saw that these are equivalent to the requirement that the chiral projections of the super Cartan curvature vanishes at the boundary which is consistent with the results obtained in \cite{Andrianopoli:2014aqa} in the non-chiral theory. In the future, it would be interesting to investigate how the definition of IH have to be extended to the supersymmetric context to rederive the boundary conditions as studied in this paper. Isolated horizons in the context of supergravity have been studied in \cite{Booth:2008ru}. There, however, one focuses on the purely bosonic sector and thus does not take fermionic degrees of freedom into account.

There are of course numerous open questions one should address in the future. For instance, this geometric approach to supergravity appears quite powerful in the appropriate description of boundaries as well as the correct implementation of locally supersymmetric boundary conditions. Moreover, as shown in \cite{Andrianopoli:2014aqa}, in this way one arrives at a very intriguing structure of the AdS supergravity Lagrangian even in case of extended supersymmetry. It would be very interesting to see how these results could be extended to higher $\mathcal{N}>2$, or even matter coupled supergravity theories. As we have demonstrated in this paper, this approach seems to be well adapted to similar questions in LQG and may shed further light on the particularity of the (graded) self-dual variables as well as their possible generalizations to extended SUGRA theories. It would also be interesting to see how the exciting recent developments \cite{Ashtekar:2020xll,Varadarajan:2021zrk} regarding the formulation of the dynamics for connection variables of gravity carries over to the supersymmetric theory. 

On the other hand, these results provide a first step toward the quantum description of boundaries in supergravity in the framework of LQG and possible applications in context of supersymmetric black holes. This requires a deeper understanding of Chern-Simons theories with supergroup as a gauge group. Super Chern-Simons theories are also of quite recent interest in context of string theory \cite{Mikhaylov:2014aoa}. As explained in the introduction, there, one observes for certain brane configurations that the boundary theory is described in terms of a super Chern-Simons theory with gauge group including e.g. the supergroups $\mathrm{OSp}(m|n)$ and $\mathrm{U}(m|n)$. We suspect that a deeper analysis of super Chern-Simons theories in the framework of LQG may also shed further light on the relation between the quantum description of boundary theories in string theory and LQG.

\section*{Acknowledgements}
We thank Abhay Ashtekar for interesting comments on the Cartan connection and the physical dimensions of the fields contained in it as well as on the role of the cosmological constant. We also would like to thank Lee Smolin for communications at an early stage of this work and in particular for his interest in application of LQG methods to extended SUGRA theories which was part of the motivation of this work. We also thank an anonymous referee for helpful comments and suggestions which helped to improve the manuscript.
K.E. thanks the German Academic Scholarship Foundation (Studienstiftung des Deutschen Volkes) for financial support. H.S. would like to acknowledge the contribution of the COST Action CA18108.

%Norbert Bodendorfer and Thomas Thiemann for helpful comments and discussions on quantizing supergravity with methods from loop quantum gravity, and Jorge Pullin and Lee Smolin for communications at an early stage of this work. We thank an anonymous referee for helpful comments and suggestions on an earlier version of this manuscript. 

\appendix

\section{Super Chern-Simons theory}\label{Chern-Simons}	
In this section, we want to briefly recall the basic definition and structure of the super Chern-Simons action. For more details on Chern-Simons theory with supergroup as a gauge group, we refer to \cite{Mikhaylov:2014aoa} as well as \cite{Cremonini:2019aao} studying the super Chern-Simons action in the geometric approach using integral forms. For more details on integral forms and related concepts see e.g. \cite{Castellani:2014goa,Catenacci:2018xsv}.\\
Before we state the super Chern-Simons action, we need to introduce invariant inner products. Let $\mathcal{G}$ be a Lie supergroup. By the super Harish-Chandra theorem, the super Lie group has the equivalent characterization in terms of a \emph{super Harish-Chandra-pair} $(G,\mathfrak{g})$ with $G$ the underlying ordinary bosonic Lie group and $\mathfrak{g}$ the super Lie algebra of $\mathfrak{g}$ with $\mathfrak{g}_{\underline{0}}=\mathrm{Lie}(G)$\footnote{For the interested reader, we note that, for sake of concreteness, we will identify the (algebro-geometric) super Lie group with the corresponding Rogers-DeWitt supergroup using the functor of points prescription (see \cite{Eder:2020erq} for more details).}.\\% Moreover, $\sigma:\,G\rightarrow\mathrm{Aut}(\mathfrak{g})$ is an extension of the Adjoint action of $G$ on $\mathfrak{g}_0$ to $\mathfrak{g}$ such that the pushforward satisfies $\sigma_*=\mathrm{ad}|_{\mathfrak{g}_0}$ \cite{}\footnote{In fact, using the functor of points of technique, it follows that one can identify $\sigma$ with the restriction of the Adjoint action $\mathrm{Ad}:\,\mathcal{G}\rightarrow\mathrm{Aut}(\mathfrak{g})$ to the bosonic subgroup $G$.}.\\
A \emph{super metric} on $\mathfrak{g}$ is a bilinear map $\braket{\cdot,\cdot}:\,\mathfrak{g}\times\mathfrak{g}\rightarrow\mathbb{C}$ that is non-degenerate and graded-symmetric, i.e. $\braket{X,Y}=(-1)^{|X||Y|}\braket{Y,X}$ for any homogeneous $X,Y\in\mathfrak{g}$. Moreover, it is called Ad-\emph{invariant}, if 
\begin{equation}
\braket{\mathrm{Ad}_gX,\mathrm{Ad}_gY}=\braket{X,Y}\quad\forall g\in G
\label{eq:A1}
\end{equation}
and 
\begin{equation}
\braket{[Z,X],Y}+(-1)^{|X||Z|}\braket{X,[Z,Y]}=0
\label{eq:A2}
\end{equation}
for all homogeneous $X,Y,Z\in\mathfrak{g}$. This can be extended to a bilinear form $\braket{\cdot\wedge\cdot}:\,\Omega^p(\mathcal{M},\mathfrak{g})\times\Omega^q(\mathcal{M},\mathfrak{g})\rightarrow\Omega^{p+q}(\mathcal{M})$ on differential forms on a supermanifold $\mathcal{M}$ with values in the super Lie algebra $\mathfrak{g}$. Therefore, first note that the sheaf $\Omega^{\bullet}(\mathcal{M},\mathfrak{g})$ carries the structure of a $\mathbb{Z}\times\mathbb{Z}_2$-bigraded module, where, for any $\omega\in(\Omega^k(\mathcal{M}))_{\underline{i}}$, its parity $\epsilon(\omega)$ is defined as 
\begin{equation}
\epsilon(\omega):=(k,\underline{i})\in\mathbb{Z}\times\mathbb{Z}_2
\label{eq:}
\end{equation}
where we will also write $|\omega|:=\underline{i}$ for the underlying $\mathbb{Z}_2$-grading. For homogeneous $\mathfrak{g}$-valued differential forms $\omega\in\Omega^p(\mathcal{M},\mathfrak{g})$ and $\eta\in\Omega^q(\mathcal{M},\mathfrak{g})$, we then set
\begin{equation}
\braket{\omega\wedge\eta}:=(-1)^{|i|(|\eta|+|j|)}\omega^i\wedge\eta^j\braket{X_i,X_j}
\label{eq:A3}
\end{equation}
where we have chosen a real homogeneous basis $(X_i)_i$ of $\mathfrak{g}$ and simply wrote $|i|:=|X_i|$ for the parity. A direct calculation yields
\begin{align}
\braket{\omega\wedge\eta}&:=(-1)^{|i|(|\eta|+|j|)}\omega^i\wedge\eta^j\braket{X_i,X_j}\nonumber\\
&=(-1)^{pq}(-1)^{|i||\eta|}(-1)^{(|\omega|+|i|)(|\eta|+|j|)}\eta^j\wedge\omega^i\braket{X_j,X_i}\nonumber\\
&=(-1)^{pq}\braket{\eta\wedge\omega}
\label{eq:A4}
\end{align}
Finally, let us derive an important identity which plays a central role in may calculations. in fact, using the $\mathrm{Ad}$-invariance (\ref{eq:A2}), one obtains
\begin{align}
\braket{\omega\wedge[\eta\wedge\xi]}&=(-1)^{|i|(|\eta|+|\xi|+|j|+|k|)}(-1)^{|j|(|\xi|+|k|)}\omega^i\wedge\eta^j\wedge\xi^k\braket{X_i,[X_j,X_k]}\nonumber\\
&=(-1)^{|i|(|\eta|+|\xi|+|j|+|k|)}(-1)^{|j|(|\xi|+|k|)}\omega^i\wedge\eta^j\wedge\xi^k\braket{[X_i,X_j],X_k}\nonumber\\
&=(-1)^{|i|(|\eta|+|j|)}\braket{\omega^i\wedge\eta^j\otimes[X_i,X_j]\wedge\xi}\nonumber\\
&=\braket{[\omega\wedge\eta]\wedge\xi}
\label{eq:A5}
\end{align}
As discussed in the main text, the Chern-Simons action naturally appears as a boundary term in chiral limit of the Holst-MacDowell-Mansouri action of supergravity. In fact, let $\mathcal{A}$ be a super connection and $F(\mathcal{A})$ its corresponding curvature, then
\begin{align}
\braket{F(\mathcal{A})\wedge F(\mathcal{A})}=\mathrm{d}\!\braket{\mathcal{A}\wedge F(\mathcal{A})-\frac{1}{6}\mathcal{A}\wedge[\mathcal{A}\wedge\mathcal{A}]}
\label{eq:A6}
\end{align}
To see this, note that
\begin{align}
\mathrm{d}\!\braket{\mathcal{A}\wedge F(\mathcal{A})-\frac{1}{6}\mathcal{A}\wedge[\mathcal{A}\wedge\mathcal{A}]}&=\braket{\mathrm{d}\mathcal{A}\wedge\mathrm{d}\mathcal{A}+\frac{1}{2}\mathrm{d}\mathcal{A}\wedge[\mathcal{A}\wedge\mathcal{A}]-\mathcal{A}\wedge[\mathrm{d}\mathcal{A}\wedge\mathcal{A}]}-\frac{1}{6}\mathrm{d}\!\braket{\mathcal{A}\wedge[\mathcal{A}\wedge \mathcal{A}]}\nonumber\\
&=\braket{\mathrm{d}\mathcal{A}\wedge\mathrm{d}\mathcal{A}+\frac{1}{3}\mathrm{d}\mathcal{A}\wedge[\mathcal{A}\wedge \mathcal{A}]-\frac{2}{3}\mathcal{A}\wedge[\mathrm{d}\mathcal{A}\wedge \mathcal{A}]}
\label{eq:A7}
\end{align}
which directly leads to \eqref{eq:A6} using $\braket{\mathcal{A}\wedge[\mathrm{d}\mathcal{A}\wedge \mathcal{A}]}=-\braket{\mathcal{A}\wedge[\mathcal{A}\wedge\mathrm{d}\mathcal{A}]}=-\braket{[\mathcal{A}\wedge \mathcal{A}]\wedge\mathrm{d}\mathcal{A}}$ which is an immediate consequence of identity \eqref{eq:A5}. When pulled back to the underlying bosonic submanifold $M$, the Chern-Simons action is thus defined as
\begin{equation}
S_{\mathrm{CS}}(\mathcal{A}):=\frac{k}{4\pi}\int_{M}{\braket{\mathcal{A}\wedge\mathrm{d}\mathcal{A}+\frac{1}{3}\mathcal{A}\wedge[\mathcal{A}\wedge\mathcal{A}]}}
\label{eq:A8}
\end{equation}
where $k$ is referred to as the \emph{level} of the Chern-Simons theory. Let us decompose $\mathcal{A}=\mathrm{pr}_{\mathfrak{g}_{\underline{0}}}\circ\mathcal{A}+\mathrm{pr}_{\mathfrak{g}_{\underline{1}}}\circ\mathcal{A}=:A+\psi$ w.r.t. the even and odd part of the super Lie algebra $\mathfrak{g}=\mathfrak{g}_{\underline{0}}\oplus\mathfrak{g}_{\underline{1}}$. Inserting this into \eqref{eq:A8}, this gives
\begin{equation}
\braket{\mathcal{A}\wedge F(\mathcal{A})}=\braket{A\wedge F(A)+\frac{1}{2}A\wedge[\psi\wedge\psi]}+\braket{\psi\wedge(\mathrm{d}\psi+[A\wedge\psi])}
\label{eq:A9}
\end{equation}
On the other hand, using $\braket{\psi\wedge[A\wedge\psi]}=\braket{\psi\wedge[\psi\wedge A]}=\braket{[\psi\wedge\psi]\wedge A}$ according to \eqref{eq:A5}, we find
\begin{align}
\braket{\mathcal{A}\wedge[\mathcal{A}\wedge\mathcal{A}]}&=\braket{A\wedge[A\wedge A]+A\wedge[\psi\wedge\psi]}+2\braket{\psi\wedge[A\wedge\psi]}\nonumber\\
&=\braket{A\wedge[A\wedge A]+A\wedge[\psi\wedge\psi]}+2\braket{A\wedge[\psi\wedge\psi]}\nonumber\\
&=\braket{A\wedge[A\wedge A]+3A\wedge[\psi\wedge\psi]}
\label{eq:A10}
\end{align}
Thus, we can rewrite \eqref{eq:A8} as follows
\begin{equation}
S_{\mathrm{CS}}(\mathcal{A})=S_{\mathrm{CS}}(A)+\frac{k}{4\pi}\int_{M}{\braket{\psi\wedge D^{(A)}\psi}}
\label{eq:A11}
\end{equation}
with $S_{\mathrm{CS}}(A)$ the Chern-Simons action of the bosonic connection $A$ and $D^{(A)}$ the associated exterior covariant derivative.

\section{Gamma-matrices and algebra}\label{Appendix:Gamma}
We summarize some important formulas concerning gamma matrices in Minkowksi spacetime and their chiral representation as used in the main text. The Clifford algebra $\mathrm{Cl}(\mathbb{R}^{1,3},\eta)$ of four-dimensional Minkowski spacetime is generated by gamma matrices $\gamma_I$, $I\in\{0,\ldots,3\}$, satisfying the Clifford algebra relations
\begin{equation}
    \left\{\gamma_{I},\gamma_{J}\right\}=2\eta_{IJ}
		\label{eq:B1.1}
\end{equation}
where the Minkowski $\eta$ is chosen with signature $\eta=\mathrm{diag}(-+++)$. In the chiral theory, we are working in the \emph{chiral} or \emph{Weyl representation} in which the gamma matrices take the form
\begin{equation}
\gamma_{I}=\begin{pmatrix}
0 & \sigma_{I}\\
\bar{\sigma}_{I} & 0
\end{pmatrix}\quad\text{and}\quad\gamma_{*}=\begin{pmatrix}
\mathds{1} & 0\\
0&-\mathds{1}
\end{pmatrix}
\label{eq:B1.2}
\end{equation}
with $\gamma_*:=i\gamma_0\gamma_1\gamma_2\gamma_3$ the highest rank Clifford algebra element also commonly denoted by $\gamma_5$ in four spacetime dimensions. Here,
\begin{equation}
\sigma_{I}=(-\mathds{1},\sigma_i)\quad\text{and}\quad\bar{\sigma}_{I}=(\mathds{1},\sigma_i)
\label{eq:B1.3}
\end{equation}
with $\sigma_i$, $i=1,2,3$ the ordinary Pauli matrices. It follows that $\mathrm{Cl}(\mathbb{R}^{s,t},\eta)$ is real vector space of dimension $\mathrm{dim}\,\mathrm{Cl}(\mathbb{R}^{s,t},\eta)=16$ spanned by the unit $\mathds{1}$ together with elements of the form
\begin{equation}
    \gamma_{I_{1}I_2\cdots I_{k}}:=\gamma_{[I_1}\gamma_{I_2}\cdot\ldots\cdot\gamma_{I_k]}
    \label{eq:B1.4.1}
\end{equation}
for $k=1,\ldots,4$, where the bracket denotes anti-symmetrization.\\
A useful formula which interrelates elements of the form (\ref{eq:B1.4.1}) with different degree is given by the following
\begin{equation}
    \tensor{\gamma}{^{I_1 I_2\ldots I_r}}\gamma_{*}=\frac{i}{(4-r)!}\tensor{\epsilon}{^{I_r I_{r-1}\ldots I_1 J_1\ldots J_{4-r}}}\tensor{\gamma}{_{J_1\ldots J_{4-r}}}
    \label{eq:B1.4}
\end{equation}
for $0\leq r\leq 4$, which will be often needed in the main text. Here, $\tensor{\epsilon}{^{IJKL}}=-\tensor{\epsilon}{_{IJKL}}$ denotes the completely antisymmetric symbol in $D=4$ with the convention $\tensor{\epsilon}{^{0123}}=1$.\\
In the context of $\mathcal{N}=2$ supergravity, we further note an important identity stemming from the Fierz-rearrangement formula. Let $\Psi_r$, for $r=1,2$, be spinor-valued one-forms. The chiral projections will be denoted $\psi_r:=\frac{1}{2}(\mathds{1}+\gamma_*)\Psi_r$ and $\psi^r:=\frac{1}{2}(\mathds{1}-\gamma_*)\Psi_r$ for $r=1,2$, respectively. One then has
\begin{align}
\psi_r\wedge\bar{\psi}_s&=\frac{1}{2}\bar{\psi}_s\wedge\psi_r-\frac{1}{8}\gamma_{IJ}\bar{\psi}_s\wedge\gamma^{IJ}\psi_r\label{eq:B1.5a}\\
\psi_r\wedge\bar{\psi}^s&=\frac{1}{2}\gamma_{I}\bar{\psi}^s\wedge\gamma^{I}\psi_r
\label{eq:B1.5b}
\end{align}
Furthermore, one has
\begin{align}
\bar{\psi}^r\wedge\psi^s\wedge\bar{\psi}_s\wedge\psi_r&=-\frac{1}{2}\bar{\psi}^r\wedge\psi^s\epsilon_{rs}\wedge\bar{\psi}_p\wedge\psi_q\epsilon^{pq}
\label{eq:B1.6}
\end{align}
Finally, in context of the chiral theory, we will often work with the spinorial index convention $\sigma_I^{AA'}$ and $\bar{\sigma}_{I A'A}$ for the Pauli-matrices. These can be used to map the internal indices $I$ of the co-frame $e^I$ to spinorial indices setting
\begin{equation}
e_{\mu}^{AA'}=e^I_{\mu}\sigma_I^{AA'}
\label{eq:B1.7}
\end{equation}
Primed and unprimed Spinor indices are raised and lowered with respect to the complete antsymmetric symbols $\epsilon^{A'B'}$ and $\epsilon^{AB}$, respectively, with the convention
\begin{equation}
\psi_A=\psi^B\epsilon_{BA}\quad\text{and}\quad\psi^A=\epsilon^{AB}\psi_B
\label{eq:B1.8}
\end{equation}
and analogously for primed indices. Due to $\epsilon\sigma_i\epsilon=\sigma_i^T$, one has the useful formula
\begin{equation}
\sigma_{I AA'}=\sigma_I^{BB'}\epsilon_{BA}\epsilon_{B'A'}=-\bar{\sigma}_{I A'A}
\label{eq:B1.9}
\end{equation}
%Using (\ref{eq:A1.3}) as well as (\ref{eq:A1.6}), it is easy to see that
%\begin{equation}
%\sigma^I_{AA'}\sigma_J^{AA'}=-2\delta_J^I
%\label{eq:A1.7}
%\end{equation}

\section{The super Poncaré and anti-de Sitter group}\label{Appendix:Supergroups}
In this section, let us briefly review the basic supergroups and algebras that play a central role in contexr of supergravity in $D=4$ spacetime dimensions (see e.g. \cite{Nicolai:1984hb,Freedman:1983na,Freedman:2012zz,Wipf:2016} for a more detailed exposition as well as \cite{Eder:2020erq} for our choice of conventions).\\
Let $\gamma^I$, $I=0,\ldots,3$, be the gamma matrices as in section \ref{Appendix:Gamma}. We then define totally antisymmetric matrices $\Xi^{AB}$, $A,B=0,\ldots,4$, via
\begin{equation}
\Xi^{IJ}:=\frac{1}{2}\gamma^{IJ}:=\frac{1}{4}[\gamma^I,\gamma^J]\quad\text{as well as}\quad\Xi^{4I}:=-\gamma^{I4}:=\frac{1}{2}\gamma^I
\label{eq:C1}
\end{equation}
where indices are raised and lowered w.r.t. the metric $\eta_{AB}=\mathrm{diag}(-+++-)$. These satisfy the following commutation relations
\begin{equation}
[\Xi_{AB},\Xi_{CD}]=\eta_{BC}\Xi_{AD}-\eta_{AC}\Xi_{BD}-\eta_{BD}\Xi_{AC}+\eta_{AD}\Xi_{BC}
\label{eq:C2}
\end{equation}
and thus provide a representation of $\mathfrak{so}(2,3)$, Lie algebra of the isometry group $\mathrm{SO}(2,3)$ of anti-de Sitter spacetime $\mathrm{AdS}_4$. Moreover, due to
\begin{equation}
(C\Xi_{AB})^T=C\Xi_{AB}
\label{eq:C3}
\end{equation}
with $C$ the charge conjugation matrix, it follows that $\Xi_{AB}$ generate $\mathfrak{sp}(4)$ the Lie algebra universal covering group $\mathrm{Sp}(4,\mathbb{R})$ of $\mathrm{SO}(2,3)$\\
The graded extension of the anti-de Sitter group with $\mathcal{N}$-fermionic generators is given by the orthosymplectic Lie group $\mathrm{OSp}(\mathcal{N}|4)$ containing $\mathrm{O}(\mathcal{N})\times\mathrm{Sp}(4)$ as a bosonic subgroup and which, on the super vector space $\mathcal{V}=(\Lambda^{\mathbb{C}})^{\mathcal{N},4}$ with $\Lambda$ a real Grassmann-algebra, is defined w.r.t. the bilinear form induced by
\begin{equation}
\Omega=\begin{pmatrix}
	\mathds{1} & 0\\
	0 & C
\end{pmatrix}
\label{eq:C4}
\end{equation}
The algebra $\mathfrak{osp}(\mathcal{N}|4)$ is then generated by all $X\in\mathfrak{gl}(\mathcal{V})$ satisfying
\begin{equation}
X^{sT}\Omega+\Omega X=0
\label{eq:C5}
\end{equation}
where $X^{sT}$ denotes the super transpose of $X$. The bosonic generators of super Lie algebra are given by
\begin{equation}
M_{AB}:=\begin{pmatrix}
	0 & 0\\
	0 & \Xi_{AB}
\end{pmatrix}\quad\text{and}\quad T^{rs}:=\begin{pmatrix}
	A^{rs} & 0\\
	0 & 0
\end{pmatrix}
\label{eq:C6}
\end{equation}
respectively, where $(A^{rs})_{pq}:=2\delta_p^{[r}\delta_q^{s]}$, $p,q,r,s=1,\ldots,\mathcal{N}$. The fermionic generators are given by
\begin{equation}
Q_{\alpha}^r:=\begin{pmatrix}
	0 & -\bar{e}_{\alpha}\otimes e_r\\
	e_{\alpha}\otimes e_r^T & 0
\end{pmatrix}
\label{eq:C7}
\end{equation}
with $(\bar{e}_{\alpha})_{\beta}=C_{\alpha\beta}$. Setting $P_I:=\frac{1}{L}\Xi_{4I}$, and rescaling $Q_{\alpha}^r\rightarrow Q_{\alpha}^r/\sqrt{2L}$ as well as $T^{rs}\rightarrow T^{rs}/2L$, one obtains the following (graded) commutation relations
\begin{align}
[M_{IJ},Q^r_{\alpha}]&=\frac{1}{2}Q_{\beta}^r\tensor{(\gamma_{IJ})}{^{\beta}_{\alpha}}\label{eq:C8a}\\
[P_I,Q^r_{\alpha}]&=-\frac{1}{2L}Q^r_{\beta}\tensor{(\gamma_I)}{^{\beta}_{\alpha}}\label{eq:C8b}\\
[T^{pq},Q_{\alpha}^r]&=\frac{1}{2L}(\delta^{qr}Q_{\alpha}^p-\delta^{pr}Q_{\alpha}^q)\label{eq:C8c}\\
[Q_{\alpha}^r,Q_{\beta}^s]=\delta^{rs}\frac{1}{2}(C\gamma^I)_{\alpha\beta}P_I+&\delta^{rs}\frac{1}{4L}(C\gamma^{IJ})_{\alpha\beta}M_{IJ}-C_{\alpha\beta}T^{rs}
\label{eq:C8d}
\end{align}
which in the limit $L\rightarrow\infty$ leads to the respective super Poincaré Lie algebra.\\
The orthosymplectic and Poincaré superalgebra contain a proper subalgebra which appears in context of chiral supergravity.  Let $T^{\pm}_i$ be defined as  
\begin{equation}
T_i^{\pm}=\frac{1}{2}(-\frac{1}{2}\tensor{\epsilon}{_{i}^{jk}}M_{jk}\pm iM_{0i})
\label{eq:C9}
\end{equation}
satisfying the commutation relations 
\begin{equation}
[T^{\pm}_i,T_j^{\pm}]=\tensor{\epsilon}{_{ij}^k}T_k^{\pm}
\label{eq:C10}
\end{equation} 
Since, the $R$-symmetry generators do not mix the chiral components of the Majorana generators $Q_{\alpha}^r$, it follows that $(T_i^+,T_{rs},Q_{A}^r)$ form a proper chiral sub super Lie algebra of $\mathfrak{osp}(\mathcal{N}|4)_{\mathbb{C}}$ with the graded commutation relations
\begin{align}
[T^+_i,T_j^+]&=\tensor{\epsilon}{_{ij}^k}T_k^+\label{eq:C11a}\\
[T_i^+,Q^r_A]&=Q^r_{B}\tensor{(\tau_i)}{^B_A}\label{eq:C11b}\\
[Q^r_{A},Q^s_{B}]&=\delta^{rs}\frac{1}{L}(\epsilon\sigma^i)_{AB}T_i^+-\frac{i}{2L}\epsilon_{AB}T^{rs}\label{eq:C11c}\\
[T^{pq},Q_{A}^{r}]&=\frac{1}{2L}(\delta^{qr}Q_{A}^p-\delta^{pr}Q_{A}^q)
\label{eq:C11d}
\end{align}
yielding the complex orthosymplectic Lie superalgebra $\mathfrak{osp}(\mathcal{N}|2)_{\mathbb{C}}$, the extended supersymmetric generalization of the isometry algebra of $D=2$ anti-de Sitter space . In the limit $L\rightarrow\infty$, this yields the extended $D=2$ super Poincaré algebra.

\end{document}